\documentclass[10pt,aps,amsfonts,amssymb,floatfix,longbibliography,superscriptaddress,notitlepage,twocolumn,footinbib]{revtex4-1}
\usepackage{mathrsfs} 
\usepackage{graphicx,color} 
\usepackage{mathtools}
\usepackage{bigints}
\usepackage{bm}
\usepackage[normalem]{ulem}
\usepackage{physics}
\usepackage{afterpage}
\usepackage{ulem}
\usepackage[colorlinks=true]{hyperref}
\hypersetup{
  colorlinks=true,
  linkcolor=blue,
  filecolor=magenta,
  citecolor=blue,
  urlcolor=blue,
}

\def\to{\rightarrow}

\newcommand{\eq}[1]{Eq.~(\ref{#1})}	

\newcommand{\Figcaption}[1]{\def\@captype{figure}\caption{#1}}
\newcommand{\tblcaption}[1]{\def\@captype{table}\caption{#1}}
\makeatother

\newcommand{\bint}{\bigintssss}
\newcounter{num}

\newcommand{\ctext}[1]{\raise0.1ex\hbox{\scriptsize \textcircled{\tiny {#1}}}}
\newcommand{\cctext}[1]{\raise0.1ex\hbox{\textcircled{\scriptsize {#1}}}}

\newcommand{\tri}{\triangle}
\newcommand{\beq}{\begin{equation}}
\newcommand{\eeq}{\end{equation}}

\begin{document}

\title{Replica theory for disorder-free spin-lattice glass transition on a tree -like simplex network}

\author{Kota Mitsumoto}
\affiliation{Institute of Industrial Science, University of Tokyo, 4-6-1 Komaba, Meguro-ku, Tokyo 153-8505, Japan}
\author{Hajime Yoshino}
\affiliation{Cybermedia Center, Osaka University, Toyonaka, Osaka 560-0043, Japan}
\affiliation{Graduate School of Science, Osaka University, Toyonaka, Osaka 560-0043, Japan}

\begin{abstract}
A class of  pyrochlore oxides, $A_2$Mo$_2$O$_7$ ($A =$ Ho, Y, Dy, Tb) with magnetic ions on corner-sharing tetrahedra
is known to exhibit spin-glass transitions without appreciable amount of quenched disorder.
Recently a disorder-free theoretical model for such a system has been
proposed which takes into account not only spins but also lattice distortions as dynamical variables
[K. Mitsumoto, C. Hotta and H. Yoshino, Phys. Rev. Lett. {\bf 124}, 087201 (2020)].
In the present paper we develop and analyze an exactly solvable disorder-free mean-field model
which is a higher-dimensional counterpart of the model.
We find the system exhibit complex free-energy landscape accompanying replica symmetry breaking
through the spin-lattice coupling.
\end{abstract}

\maketitle

\section{Introduction}

Glass formation is a generic phenomenon that occurs in various systems, including structural glasses \cite{berthier2011theoretical, tarjus2011overview, parisi2020theory}, spin glasses \cite{mezard1987spin, mydosh1993spin, kawamura2015spin}, orbital glasses \cite{fichtl2005orbital, thygesen2017orbital, mitsumoto2020spin}, and charge glasses \cite{kagawa2013charge}.
While the possibility of genuine thermodynamic glass transition is highly debated in the case of structural glasses\cite{berthier2011theoretical}, thermodynamic spin glass transitions in magnetic systems with strong quenched disorder is well established experimentally \cite{mydosh1993spin}. The thermodynamic spin-glass transition is most unambiguously demonstrated by the measurements of the negatively diverging non-linear susceptibility with critical scaling at the spin-glass transition temperature \cite{chikazawa1980nonlinear,taniguchi1985nonlinear,levy1986nonlinear}. On the theoretical side, Edwards and Anderson proposed that the origin of the spin glass with quenched disorder is randomness and frustration \cite{edwards1975theory}, which is now widely accepted through analysis of the Edwards-Anderson (EA) model by mean-field theories exact in the large dimensional limit  \cite{sherrington1975solvable,almeida1978stability, parisi1979infinite}
and by numerical simulations at 3 dimensions on both Ising \cite{ogielski1985dynamics, bhatt1988numerical, kawashima1996phase} and Heisenberg \cite{hukushima2000chiral, lee2007large, ogawa2020monte} spins. Remarkably thermodynamic spin-glass transitions have been established experimentally also in a class of systems {\it without quenched disorder}, pyrochlore antiferromagnets \cite{greedan1986spin, gaulin1992spin, dunsiger1996muon, gingras1997static, gardner1999glassy, hanasaki2007nature, gardner2010magnetic, thygesen2017orbital, mitsumoto2020spin} through experiments including, in particular, the measurements of the negatively diverging non-linear susceptibility \cite{gingras1997static}. On the theoretical side, the explanation for the sharp spin-glass transition without quenched disorder remained a challenging open problem.

\begin{figure}[t]
\centering
\includegraphics[width=85mm]{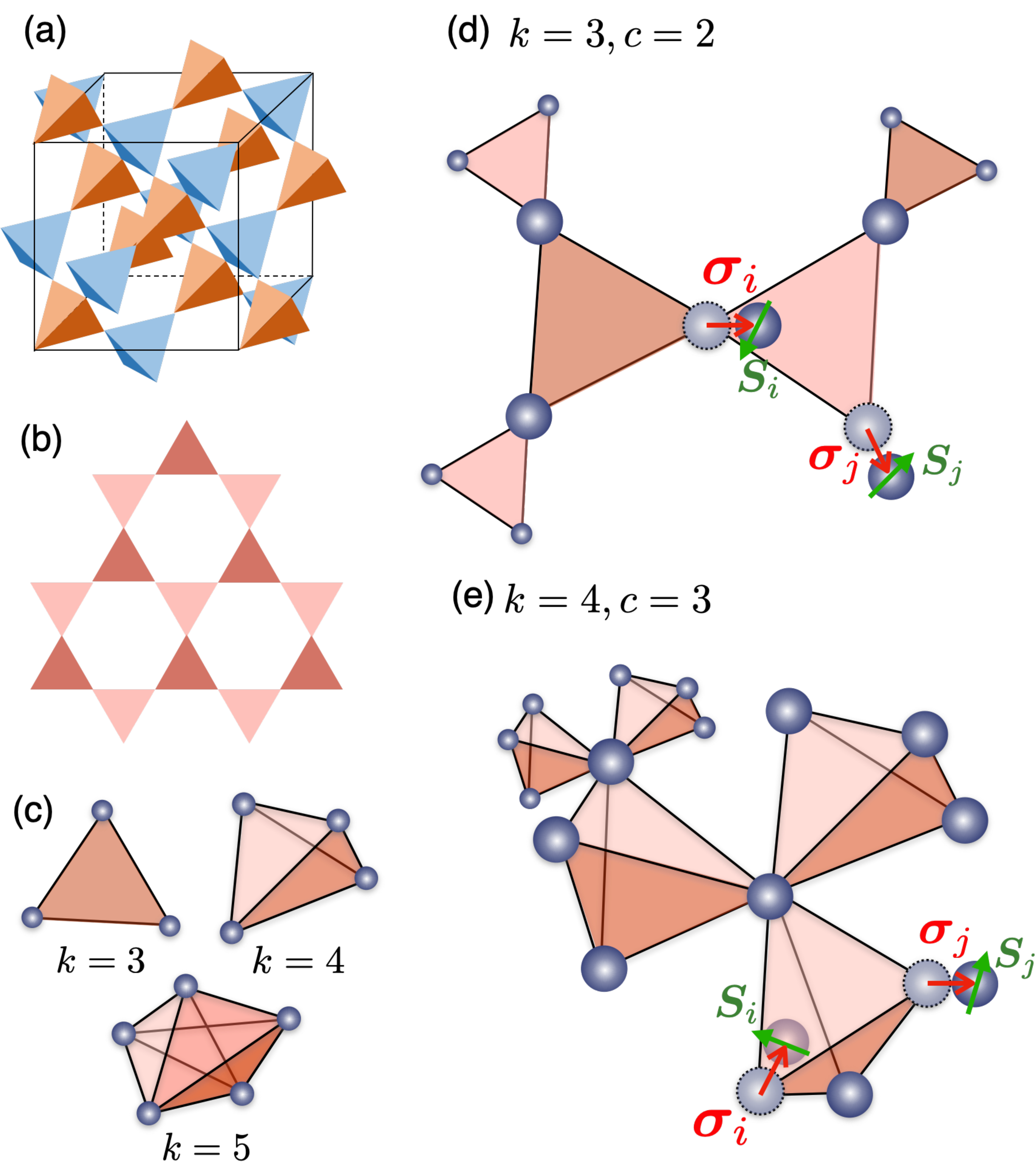}
\caption{(a) Pyrochlore lattice and (b) kagom\'{e} lattice.
(c) $(k-1)$-dimensional simplices in the cases of $k=3,4$ and 5.
(d, e) The corner-sharing ($k-1$)-simplices network in the cases of (d) $k=3$ and $c=2$ and (e) $k=4$ and $c=3$.
}
\label{simplex_network}
\end{figure}

A common feature in the glassy systems mentioned above is frustration: strong many-body interactions competing with each other. The resultant glassiness is often viewed in a phenomenological complex free-energy landscape picture: free-energy landscape with multiple nearly degenerate minima separated by high free-energy barriers. One important aspect of such a free-energy landscape is the near degeneracy of the energy minima or {\it flatness}. Frustration can realize such flat energy landscapes. Indeed, as a primary approximation, the pyrochlore antiferromagnets can be modeled by the classical Heisenberg spin model with the nearest-neighbor antiferromagnetic interaction on the pyrochlore lattice which is a three-dimensional corner-sharing network of tetrahedra (Fig. \ref{simplex_network} (a)). Because of the extremely strong geometrical frustration, the low energy landscape of the model exhibits flatness and the system remains in a classical spin liquid state down to $T=0$ \cite{reimers1991mean, moessner1998properties, gardner2010magnetic}. However, apparently, something is missing in this simplest model as it fails to capture the spin-glass transition found experimentally. In short, it does not realize high free-energy barriers indispensable for glassiness. In the case of spin glasses with strong quenched disorder, the energy landscape is distorted randomly by the quenched disorder leading to high free-energy barriers.
Then a simple solution is to add some quenched disorder onto the antiferromagnetic model  \cite{saunders2007spin, shinaoka2011spin} but such a disorder has never been observed in the pyrochlore oxides. How a lugged free-energy landscape with high barriers can be created by distorting a flat energy landscape without the help of such quenched disorder is a non-trivial problem.

Recently, a microscopic theoretical model \cite{mitsumoto2020spin} aimed to capture the emergence of disorder-free spin-glass transition on the pyrochlore oxides was proposed. It was argued that the model realizes a complex free-energy landscape by coupling two distinct dynamical variables which are strongly frustrated among themselves. One is the set of classical Heisenberg spins on the vertices (molybdenum ions) which are interacting anti-ferromagnetically with the nearest neighbors. As just mentioned above this anti-ferromagnetic spin system on the pyrochlore lattice is highly frustrated giving rise to the flat energy landscape.
The other degree of freedom
is the set of spatial displacements of the vertices, the molybdenum ions.
Following the suggestion by a recent experiment \cite{thygesen2017orbital},
they are assumed to be displaced either towards ($in$) or away from ($out$) the center of the tetrahedron following two-$in$-two-$out$ ice rule \cite{pauling1935structure} at each tetrahedron. This is equivalent to the so-called `spin ice'
which is another highly frustrated system on the pyrochlore lattice with a flat energy landscape (only low enough barriers) that allows liquid state down to $T=0$ \cite{gardner2010magnetic, bramwell2020history, udagawa2021spin}. In the model \cite{mitsumoto2020spin} these two distinct degrees of freedom are coupled through the Goodenough-Kanamori mechanism \cite{goodenough1955theory, kanamori1959superexchange, solovyev2003effects, smerald2019giant, mitsumoto2020spin}.
Since the two degrees of freedom are totally unrelated to each other in the absence of the coupling, the coupling can distort their flat energy landscapes.
Extensive Monte Carlo simulation \cite{mitsumoto2020spin} in the presence of the coupling suggested a glass transition where both the spins and lattice displacements become cooperatively frozen simultaneously.
This was demonstrated by two key observations.
One is the critical slowing down of the two degrees of freedom approaching
a common freezing temperature $T_{\rm c}$.
The other is an observation that
the nonlinear magnetic and dielectric susceptibilities, which are
associated with the freezing of the spins and lattice displacements respectively,
grow negatively large in the vicinity of $T_{\rm c}$.

\begin{figure*}[t]
\centering
\includegraphics[width=150mm]{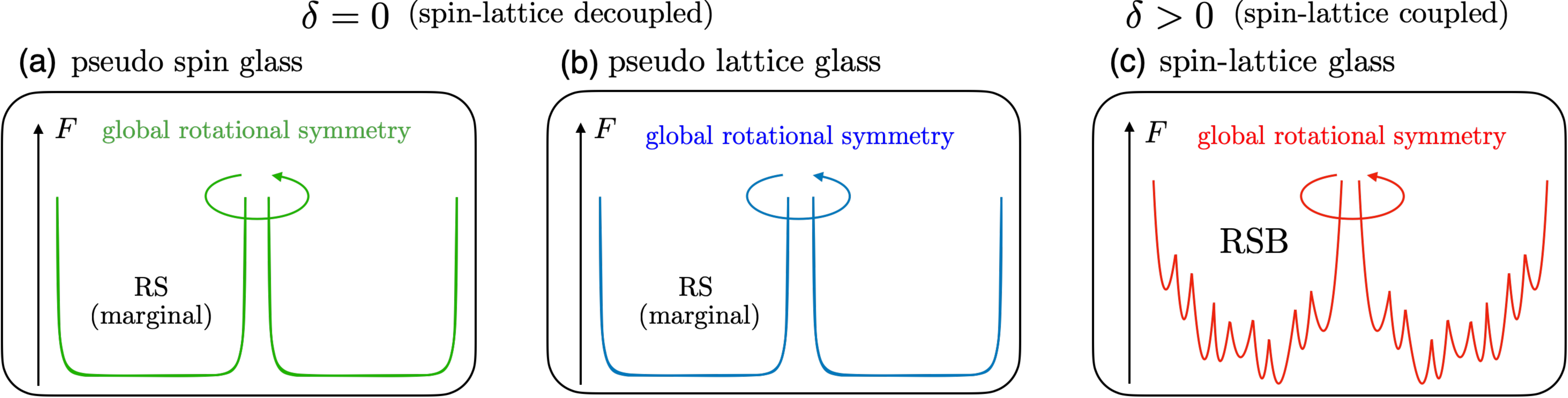}
\caption{Schematic picture of the free-energy landscape
  of the mean-field theory. (a, b) In the absence of the spin-lattice coupling $(\delta=0)$,
  the (a) spin and (b) lattice degrees of freedom independently exhibit low temperature phases
  where global rotational symmetries are broken.
  They are characterized by marginally stable replica symmetric solutions which imply
  flat landscapes of their own.
  (c) Switching on the spin-lattice coupling $(\delta>0)$,
  at low enough temperatures, the system exhibits a glass phase where both the
  spin and lattice degrees of freedom are frozen. The replica symmetry is broken there
  implying a complex free-energy landscape.
}
\label{free_energy_landscape}
\end{figure*}

To obtain further insights into the problem, we develop an exactly solvable, disorder-free mean-field model which can be regarded as a variant of the model \cite{mitsumoto2020spin}
in a high-dimensional limit $d \to \infty$ in the present paper.
We analyze the model using the replica method.
Recently the replica approach, which was originally developed for systems with quenched disorder \cite{edwards1975theory, sherrington1975solvable, parisi1979infinite}, was applied successfully to describe glassy systems without quenched disorder \cite{charbonneau2014fractal,yoshino2018disorder}. In the nutshell, one considers the effect of infinitesimal random pinning field which is
switched off after taking the thermodynamic limit $N \to \infty$ \cite{parisi1989mechanism,monasson1995structural,yoshino2023lecture}. The replica approach established existence of genuine thermodynamic glass transitions in high-dimensional limits
in the absence of quenched disorder:
structural glasses \cite{charbonneau2014fractal} and disorder-free $p$-body interacting spin systems \cite{yoshino2018disorder}.
The mean-field approach is useful to elucidate relevant order parameters,
which are overlaps of dynamical degrees of freedom belonging to different replicas in the case of glasses, and to extract mean-field phase diagrams, which serves as useful guidelines.
It is particularly helpful in understanding the formation of glasses
\cite{MP99, parisi2020theory}
as it derives microscopically the complex free-energy landscape, which has been anticipated phenomenologically,
based on the notion of replica symmetry breaking (RSB) \cite{almeida1978stability, parisi1979infinite,mezard1987spin}. On the other hand, one should always keep in mind that the mean-field approaches are exact only at high enough dimensions. In general phase transitions found at high dimensions disappear at low enough dimensions because of thermal fluctuations. More seriously some of the predictions made by mean-field theory can be problematic
such that they only hold strictly in high dimensional limits \cite{fisher1991pathologies}.

We define our mean-field model for the spin-lattice coupled systems on a tree-like simplex lattice \cite{husimi1950note, RevModPhys.42.271, rieger1992disordered, chandra1994spin, udagawa2019spectrum, cugliandolo2020mean} which is a higher-dimensional extension of the pyrochlore lattice and kagom\'{e} lattice (Figs. \ref{simplex_network} (a) and (b)), and solve it exactly by the replica method in a dense connectivity limit corresponding to the large dimensional limit. Here, a simplex means a geometrical unit in which all vertices are fully connected, and a simplex with $k$ vertices is called a $(k-1)$-dimensional simplex. 
For example, triangle, tetrahedron and $5$-cell are $(k-1)$-dimensional simplex with $k=3, 4$ and $5$, respectively, as shown in Fig. \ref{simplex_network} (c). For simplicity, both the spins and lattice displacements are modeled by
continuous variables subjected to spherical constraints. We derive exact free-energy functional in terms of glass
order parameters for the spins and lattice displacements following the approach of \cite{yoshino2018disorder}. Then we analyze the problem within the replica symmetric (RS) ansatz and examine the stability of the solution to detect instability toward RSB.

In Fig.~\ref{free_energy_landscape}, we show the schematic free-energy landscape predicted by the theory. In the absence of the spin-lattice coupling by the
Goodenough-Kanamori mechanism $(\delta=0)$ (Figs.~\ref{free_energy_landscape} (a) and (b)), the spin and lattice degrees of freedom independently exhibit low temperature phases where their global rotational symmetries are broken.
The low temperature phases
are characterized by marginally stable replica symmetric solutions which imply
flat landscapes of their own. The independent flat landscapes may be related to the
those of the corresponding 3 dimensional systems: the antiferromagnet model
\cite{reimers1991mean, moessner1998properties, gardner2010magnetic}
and the spin-ice system \cite{gardner2010magnetic, bramwell2020history, udagawa2021spin}
on the pyrochlore lattice. However, we believe that the breaking of the global rotational symmetry
is an artifact in the high dimensional limit.
In the presence of the coupling $(\delta >0)$ (Fig.~\ref{free_energy_landscape} (c)), we find a glass phase
at low enough temperatures
where both spins and lattice degrees of freedom are frozen cooperatively.
We find the replica symmetric solution is unstable there suggesting
spontaneous replica symmetry breaking (RSB). This strongly implies the emergence of
a complex free-energy landscape.

The rest of this paper is organized as follows.
In Sec. \ref{sec:model}, we introduce our mean-field model.
We present the development of the replica theory for our model in Sec. \ref{sec:replica}.
In Sec. \ref{sec:result}, we present the analysis of the replica theory based on the replica symmetric (RS) ansatz:
the phase diagram  (Sec. \ref{sec:phase}) including the spin-lattice decoupled case (Sec. \ref{sec:decoupled}), the analysis of the stability condition of RS ansatz, the asymptotic behavior of the glass susceptibilities near the glass transition temperatures, and the behavior of the internal energy and the heat capacity for both separated (Sec. \ref{sec:separated}) and simultaneous (Sec. \ref{sec:simultaneous}) glass transition cases.
A summary and discussion of the results are given in Sec. \ref{sec:summary}, including a comparison with the three-dimensional pyrochlore system.
The details of the derivation of the free-energy functional, the analysis of the stability condition, and the computation of glass susceptibility are presented in Appendix. \ref{sec:derivation}, \ref{sec:stability} and \ref{sec:sus}, respectively. Finally in Appendix. \ref{sec:ene} the definition of the internal energy and the heat capacity with the RS ansatz are presented.

\section{Model}\label{sec:model}

\subsection{Tree-like simplex network}
\label{sec-tree-like-simplex-network}

Our model consists of classical spins and lattice distortions on a corner-sharing $(k-1)$-dimensional simplices network with connectivity $c$ which is the number of tetrahedra sharing one lattice site.
The pyrochlore lattice and kagom\'{e} lattice correspond to the cases of $(c,k) = (2,4)$ and $(c,k) = (2,3)$, respectively, as shown in Figs. \ref{simplex_network} (a) and (b).
In order to develop an exactly solvable mean-field model, we consider a tree-like graph of simplices with connectivity $c$, as shown in Figs. \ref{simplex_network} (d) and (e) so that the presence of loops can be neglected except for the local triangles within the simplices.  More precisely we consider dense limit $N \gg c \gg 1$\cite{yoshino2018disorder} which greatly simplifies the theory. 


\subsection{Spin-lattice coupled model}\label{sec:slmodel}

We assume that spins are $M$-component classical vector spins $\bm{S}_i = (S_i^1, S_i^2, ..., S_i^M)$ $(i=1,2,...,N)$ subjected to a spherical constraint $\sum_{i=1}^N|\bm{S}_i|^2=NM$.
We assume that the lattice distortions are $c$-component vectors $\bm{\sigma} = (\sigma_i^1, \sigma_i^2,...,\sigma_i^c)$ subjected to a spherical constraint $\sum_{i=1}^N|\bm{\sigma}_i|^2 = Nc$.
Here $c$ is the connectivity of the lattice which plays the role of spacial dimension within our model.
In order to obtain a non-trivial theory in the dense limit $N \gg c \gg 1$,
we also assume $M  \gg 1$ with,
\beq
\alpha = \frac{c}{M}
\label{eq-def-alpha}
\eeq
being fixed to a value of $O(1)$.
%

The Hamiltonian is given by,
\begin{align}
H &= \frac{1}{\sqrt{c(k-1)}}\sum_{\langle i,j \rangle} \qty[ J_{\sigma_i, \sigma_j} \bm{S}_i \cdot \bm{S}_j + \epsilon \bm{\sigma}_i \cdot \bm{\sigma}_j], \label{hami1} \\
J_{\sigma_i, \sigma_j} &= J[1+ \delta (\bm{\sigma}_i \cdot \hat{\bm{r}}_{ij} + \bm{\sigma}_j \cdot \hat{\bm{r}}_{ji})],~~~(J>0)
\label{hami1b}
\end{align}
where $\langle i,j \rangle$ denotes the summation over the nearest-neighboring pairs, $\hat{\bm{r}}_{ij}$ is the unit vector from the $i$-th site to the $j$-th site, $J$ and $\epsilon$ are energy scales of spin-spin and lattice-lattice interactions, respectively. The parameter $\epsilon$ reflects the elastic energy cost of lattice distortion \cite{mitsumoto2022supercooled} so that we call it rigidity in the following.
The spin-exchange interaction is antiferromagnetic without lattice distortion and is modified by the lattice distortion via the Goodenough-Kanamori mechanism \cite{goodenough1955theory, kanamori1959superexchange}. The parameter $\delta$ controls the strength of spin-lattice coupling.

Let us note that the Hamiltonian is invariant under the global reflection of the spins
$\bm{S}_{i} \to -\bm{S}_{i}$ for $\forall i$. On the other hand, it is not invariant under
the global reflection of the lattice distortions $\bm{\sigma}_{i} \to -\bm{\sigma}_{i}$ for $\forall i$  in general except in the absence of the spin-lattice coupling $\delta=0$.

The present model can be regarded as a high dimensional variant of the model defined in Ref. \cite{mitsumoto2020spin} for the pyrochlore system. In the pyrochlore system, if the spin-lattice coupling is switched off $\delta=0$, both spin and lattice distortion remain in liquid (paramagnetic) phase down to $T=0$ due to frustration. In the presence of the coupling $\delta > 0$, if the energy scale of lattice-lattice interactions is not too small, numerical simulations suggest the system exhibits a simultaneous spin and lattice glass transition at a critical temperature \cite{mitsumoto2020spin}.

On the corner-sharing $(k-1)$-simplices network, we can formally rewrite the Hamiltonian \eq{hami1} as the summation of the potential energy running over all simplices $\tri=1,2,\ldots, N_{\tri}$,
\beq
H = \sum_{\tri = 1}^{N_{\tri}}V(\vec{\bm{S}}_{(\tri)},\vec{\sigma}_{(\tri)})
\label{hami2}
\eeq
with 
\begin{align}
\nonumber
  &V(\vec{\bm{S}}_{(\tri)}, \vec{\sigma}_{(\tri)}) \\
  &= \frac{1}{\sqrt{c(k-1)}}\sum_{\langle i,j \rangle \in \tri} [J_{\sigma_{i},\sigma_{j}}\bm{S}_{i}\cdot\bm{S}_{j} +\epsilon \bm{\sigma}_{i} \cdot \bm{\sigma}_{j} ],
  \label{hami2b}
\end{align}
where
\beq
N_{\tri} = N\frac{c}{k}
\label{eq-N_tri}
\eeq
is the number of simplices in the whole system, $\vec{\bm{S}}_{(\tri)} = (\bm{S}_{1(\tri)},\bm{S}_{2(\tri)},...,\bm{S}_{k(\tri)})$ and $\vec{\bm{\sigma}}_{(\tri)} = (\bm{\sigma}_{1(\tri)}, \bm{\sigma}_{2(\tri)},...,\bm{\sigma}_{k(\tri)})$ represent the set of spin variables and lattice displacements belonging to a given simplex $\tri$. 
The summation $\sum_{\langle i,j \rangle \in \tri}$ represents summation over nearest neighbor pairs within a simplex $\tri$.

The free energy of the system is given by $-\beta F = \log Z$, where $\beta$ is the inverse temperature and $Z$ represents partition function defined as, $Z = \text{Tr}_{\bm{S}} \text{Tr}_{\bm{\sigma}} e^{-\beta H}$,
where $\text{Tr}_{\bm{S}}$ and $\text{Tr}_{\bm{\sigma}}$ represent traces over the spin space of the spin $\bm{S}_i$ and the lattice displacement $\bm{\sigma}_i$ for all $i$ with spherical constraint,
\begin{align}
\text{Tr}_{\bm{S}} &= \prod_{i = 1}^N \prod_{\mu = 1}^M \qty(\bint_{-\infty}^{\infty} dS_i^\mu) \delta \qty(1-\frac{1}{NM}\sum_{i = 1}^N |\bm{S}_i|^2), \label{trace1} \\
\text{Tr}_{\bm{\sigma}} &= \prod_{i = 1}^N \prod_{\nu = 1}^c \qty(\bint_{-\infty}^{\infty} d\sigma_i^\nu)  \delta \qty(1- \frac{1}{Nc}\sum_{i = 1}^N |\bm{\sigma}_i|^2). \label{trace2}
\end{align}


\subsection{Ground state manifold in the absence of spin-lattice coupling}
\label{sec-ground-state-manifold-before-coupling}

In the absence of the coupling $\delta=0$,
we have $J_{\sigma_{i},\sigma_{j}} \to J$ so that
the system decouples into
two independent subsystems (see \eq{hami1}, \eq{hami1b}).
Within each of the subsystem the vectorial variables
simplex network interact with each other anti-ferromagnetically.
In the other representation of the hamiltonian
\eq{hami2} with \eq{hami2b} we find,
\begin{align}
\nonumber
&\sum_{\tri=1}^{N_{\tri}} \sqrt{c(k-1)}V(\vec{\bm{S}_{\tri}}, \vec{\sigma}_{\tri})  \\ \nonumber
\xrightarrow{\delta\to 0} &\sum_{\tri=1}^{N_{\tri}}\qty[\frac{J}{2}\left(\sum_{i \in \tri}\bm{S}_{i}\right)^{2}
+\frac{\epsilon}{2}\left(\sum_{i \in \tri}\bm{\sigma}_{i}\right)^{2}] \\
&-\frac{Nc}{2}(J M+\epsilon c),
\end{align}
where we used $\sum_i|\bm{S}_i|^{2}=NM$ and  $\sum_i|\bm{\sigma}_i|^{2}=Nc$. Thus the local Hamiltonian can be minimized
by requiring
\beq
\sum_{i \in \tri}\bm{S}_{i}=0 \qquad
\sum_{i \in \tri}\bm{\sigma}_{i}=0.
\label{eq-sum-rules}
\eeq
Furthermore, since we are considering tree-like network of simplices, 
the configurations which satisfy the local sum rules \eq{eq-sum-rules} in all simplices $\tri$ are
the ground states of the total system. 
Then a Maxwellian counting argument \cite{reimers1991mean,moessner1998properties}
can be made for the ground state manifold.
For the spin sector, we have $NM$ degrees of freedom (disregarding the small correction due to the spin normalization condition) while the sum rule \eq{eq-sum-rules} imposes $N_{\tri}M=NM(c/k)$ conditions. Then $NM(1-c/k)$ degrees of freedom remain in the ground state manifold of spins.
Similarly, $\alpha NM(1-c/k)$ degrees of freedom remain in the ground state manifold of
of the lattice distortions. Thus if we have,
\beq
0 < \frac{c}{k} < 1
\eeq
there is a macroscopic degeneracy of the ground states in the absence of spin-lattice coupling.

\subsection{Microscopic background of the model}\label{sec:background_of_model}

The microscopic background of the spin-lattice coupled model is explained in \cite{mitsumoto2020spin}.
Let us summarize it here and add some further remarks.
The lattice distortion is a dynamical Jahn-Teller distortion caused by the orbital degeneracy of the molybdenum ions \cite{mitsumoto2022supercooled} and favors 2-$in$-2-$out$ patterns following the ice rule.
The lattice distortion changes the angle between an oxygen ion O$^{2-}$ and two magnetic ions Mo$^{4+}$, modifying the superexchange interaction between spins by the Goodenough-Kanamori mechanism \cite{goodenough1955theory, kanamori1959superexchange, solovyev2003effects, smerald2019giant, mitsumoto2020spin}.
Our numerical model on a pyrochlore lattice incorporates the above two features: lattice distortions favor the ice rule, and the interaction between classical Heisenberg spins is modified by the lattice distortion patterns as antiferromagnetic for $in$-$in$ and $in$-$out$ bonds and ferromagnetic for $out$-$out$ bond.

In Ref. \cite{mitsumoto2020spin}, the dynamical Jahn-Teller effect was modeled phenomenologically
by a simple spin-ice Hamiltonian with only nearest-neighbor interactions between lattice distortions.
This was motivated by the experiment \cite{thygesen2017orbital} which 
suggested 2-$in$-2-$out$ patterns of the lattice distortions.
However in Ref. \cite{mitsumoto2022supercooled}, it was found based on a detailed
microscopic analysis of the Jahn-Teller effect, that the lattice model should be extended
to include also second and third nearest-neighbor interactions. 
It was found that the ground state of this extended lattice model obeys the 2-$in$-2-$out$
rule but it exhibits a specific periodically alternating pattern, i.~e. a long-ranged order.
However, quite surprisingly, numerical simulations of the extended model showed that the system
avoids the ordering transition even under very slow cooling
and remains in a supercooled, disordered ice state down to the lowest temperature.
This result suggests that we can assume an ice-like liquid state for the lattice degrees of freedom.
Thus for the sake of simplicity, we just assume the nearest neighbor interactions 
like the simple spin-ice Hamiltonian for the lattice distortions
and do not include further neighbor interactions in the present mean-field model. 

\section{Replica theory}\label{sec:replica}

\subsection{Free-energy functional}\label{sec:replicated}

In this section, we introduce the replica method, derive the replicated free-energy
as a functional of the order parameters and obtain the replica symmetric (RS) solution.
The details of the derivation of the free-energy functional are given in Appendix. \ref{sec:derivation}.

The free energy of the system can be obtained formally as,
\beq
-\beta F = \log Z = \lim_{n\to0} \partial_n Z^n.
\eeq
where $Z^{n}$ is the partition function of the replicated system $a=1,2,\ldots,n$,
\beq
Z^n = \prod_a \qty(\text{Tr}_{\bm{S}^a}\text{Tr}_{\bm{\sigma}^a})
e^{- \beta H_{n}}
\label{replicated}
\eeq
Here $H_{n}$ is the Hamiltonian
of the replicated system,
\beq
\beta H_{n} = \sum_{a = 1}^n \sum_{\tri} \beta V(\vec{\bm{S}}_{(\tri)}^a, \vec{\sigma}_{(\tri)}^a).
\eeq

To investigate the glass transition, we introduce the glass order parameter matrices of spins and lattice distortions as overlaps between different replicas $a$ and $b$,
\begin{align}
Q_{ab} &= \lim_{N \to \infty} \frac{1}{NM} \sum_{i=1}^N \expval{\bm{S}^a_i \cdot \bm{S}^b_i}, \label{orderQ} \\ 
q_{ab} &= \lim_{N \to \infty} \frac{1}{Nc} \sum_{i=1}^N \expval{\bm{\sigma}^a_i \cdot \bm{\sigma}^b_i}. \label{orderq}
\end{align}
For convenience, we extend the matrices to include the diagonal elements, $Q_{aa} = q_{aa} = 1$,
to take into account the spherical constraints.

As explained in Appendix. \ref{sec:derivation}, the replicated partition function can be
expressed in the dense limit $N \gg c \gg 1$ as,
\beq
Z^n \propto \prod_{a \le b} \qty( \int_{-\infty}^\infty dQ_{ab} \int_{-\infty}^\infty dq_{ab}) e^{(N c) s_n[\hat{Q}, \hat{q}]},
\eeq
where we defined,
\beq
s_n[\hat{Q}, \hat{q}] = s_{\rm ent}^{(s)}[\hat{Q}] + s_{\rm ent}^{(\sigma)}[\hat{q}] + \mathcal{F}_{\rm int}[\hat{Q},\hat{q}].
\label{Sn}
\eeq
with
\begin{align}
s_{\rm ent}^{(s)}[\hat{Q}] &= \frac{1}{\alpha} \qty[\frac{n}{2}(1+\log(2\pi)) + \frac{1}{2}\log(\det\hat{Q})], \label{entQ} \\
s_{\rm ent}^{(\sigma)}[\hat{q}] &= \frac{n}{2}(1+\log(2\pi)) + \frac{1}{2}\log (\text{det}\hat{q}), \label{entq} \\
\mathcal{F}_{\rm int}[\hat{Q}, \hat{q}] &= \frac{\beta^2}{4} \sum_{a,b} \qty[ \frac{J^2}{\alpha} \qty(1+2\delta^2q_{ab}) Q_{ab}^2 + \epsilon^2 q_{ab}^2 ]. \label{Fint}
\end{align}
Note that the partition function doesn't depend on the simplex dimension $k$
thanks to the scaling factor $1/\sqrt{k-1}$ we introduced in the Hamiltonian.
The resultant free energy functional becomes,
 \beq
 -\beta f[\hat{Q}^*, \hat{q}^*] = \frac{-\beta F[\hat{Q}^*, \hat{q}^*]}{Nc} = \eval{\pdv{s_n[\hat{Q}^*, \hat{q}^*]}{n}}_{n=0}
 \label{eq-free-energy}
 \eeq
where  $Q^{*}$ and $q^{*}$ are the saddle points which verify 
the saddle point equations,
\begin{align}
  0 = \eval{ \pdv{s_n[\hat{Q},\hat{q}]}{Q_{ab}} }_{\hat{Q} = \hat{Q}^*, \hat{q} = \hat{q}^*} = \eval{ \pdv{s_n[\hat{Q},\hat{q}]}{q_{ab}} }_{\hat{Q} = \hat{Q}^*, \hat{q} = \hat{q}^*}.
  \label{eq-SP}
\end{align}

The stability condition is also required, namely the eigenvalues of the $n(n-1) \times n(n-1)$ Hessian matrix,
\beq
\hat{\mathcal{H}} = \mqty[H_{QQ}&H_{Qq}\\H_{qQ}&H_{qq}],
\eeq 
where the components of the $\frac{n(n-1)}{2} \times \frac{n(n-1)}{2}$ block matrices $H_{QQ}, H_{Qq}=H_{qQ}$ and $H_{qq}$ are defined as
\begin{align}
H_{Q_{ab},Q_{cd}} &\equiv - \pdv{s_n[\hat{Q},\hat{q}]}{Q_{ab}}{Q_{cd}}, ~~H_{Q_{ab},q_{cd}} \equiv - \pdv{s_n[\hat{Q},\hat{q}]}{Q_{ab}}{q_{cd}}, \nonumber \\
H_{q_{ab},q_{cd}} &\equiv - \pdv{s_n[\hat{Q},\hat{q}]}{q_{ab}}{q_{cd}} \label{hessian}
\end{align}
with $a<b,c<d$, must be non-negative in the $n \to 0$ limit.
The glass susceptibility, which is associated with the nonlinear magnetic or dielectric susceptibilities, is given by the inverse matrix of the Hessian matrix,
\beq
\hat{\chi} = \mqty[\chi_{QQ} & \chi_{Qq} \\ \chi_{qQ} & \chi_{qq}] = \mqty[H_{QQ} & H_{Qq} \\ H_{qQ} & H_{qq}]^{-1} = \hat{\mathcal{H}}^{-1}.
\label{chi_sg}
\eeq
Hereafter, we call $\chi_{QQ}, \chi_{qq}$ and $\chi_{Qq}$ as the spin glass susceptibility, the lattice glass susceptibility, the cross glass susceptibility, respectively.

\subsection{Spin-lattice decoupled case}\label{sec:decoupled}

In the absence of the spin-lattice coupling, i.e $\delta=0$, the free-energy
(see  \eq{Sn}-\eq{eq-free-energy}) become decoupled into the spin part and lattice part.
The free energy associated with them is essentially the same as that of the
spherical Sherrington-Kirkpatrick (SK) model \cite{kosterlitz1976spherical}. It is known that
the latter model has a low temperature phase where the global rotational symmetry is broken.
However, the replica symmetry (RS) survives marginally in the sense that the
stability of the RS solution (see sec~\ref{sec:rs}) is marginally stable.

Thus in the absence of the spin-lattice coupling, the spin and lattice degrees of freedom
exhibit 'pseudo glass transitions' accompanying breaking of the global rotational symmetry
but without RSB. The marginally stable RS solutions imply that the free-energy
landscapes in these low temperature phases are
flat as shown schematically in Fig.~\ref{free_energy_landscape} (a) and (b).

\subsection{Replica symmetric ansatz}\label{sec:rs}

Now, our task is to solve the above problem with the simplest ansatz called the replica symmetric (RS) ansatz.
 In the RS ansatz we assume that the overlap matrices $Q_{ab}$ and $q_{ab}$ are parametrized as,
 \beq
 \begin{split}
 Q_{ab} &= (1-Q)\delta_{ab} + Q, \\
 q_{ab} &= (1-q)\delta_{ab} + q,
 \end{split}
 \label{RSansatz}
 \eeq
 where $\delta_{ab}$ is the Kronecker's delta.
Then the free energy within RS ansatz is obtained as,
\begin{align}
\nonumber
-\beta f^{\rm RS} &= \frac{1}{2\alpha}\qty(\frac{Q}{1-Q}+ \log(1-Q)) \\ \nonumber
&+ \frac{1}{2}\qty(\frac{q}{1-q} + \log(1-q)) \\ \nonumber
&+ \frac{\beta^2}{4}\Biggl\{\frac{J^2}{\alpha}(1+2\delta^2) + \epsilon^2 \\
&- \qty[\frac{J^2}{\alpha}(1+2\delta^2q)Q^2 + \epsilon^2q^2]\Biggr\} + \text{const.}
\label{RSansatz-free-energy}
\end{align}
The saddle point equations with respect to the order parameters $Q$ and $q$ are obtained as,
\begin{align}
0 & = Q\qty[\frac{1}{(1-Q)^2} -  (\beta J)^2(1+2\delta^2 q)]
\label{RSsaddleQ1} \\
0 & =  \frac{q}{(1-q)^2} - ( \beta \epsilon)^2q - \frac{(\beta J \delta)^2}{\alpha}Q^2.
\label{RSsaddleq1}
\end{align}
We see that $Q=q = 0$ is always a solution which represents
the high temperature disordered (paramagnetic or liquid) state.
Below the glass transition temperatures (See Appendix. \ref{sec:stability}) this solution becomes unstable. 

\section{Results}\label{sec:result}

\subsection{Phase diagram}\label{sec:phase}

\begin{figure*}[t]
\centering
\includegraphics[width=0.85\textwidth]{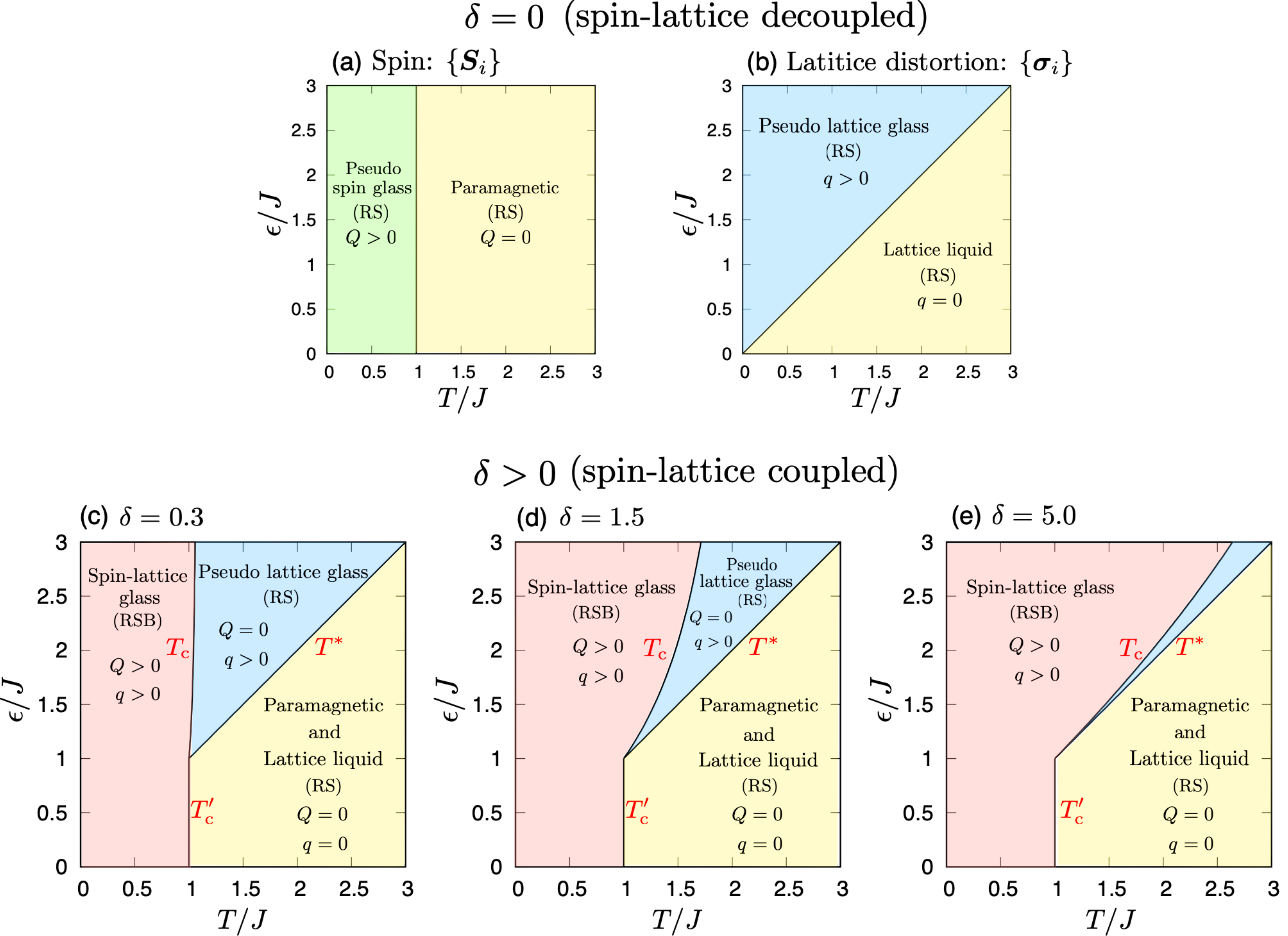}
\caption{
Mean-field phase diagrams for $\delta=0$ and $\delta > 0$.
}
\label{phase_mean}
\end{figure*}

We show in Fig. \ref{phase_mean} the typical phase diagram, which is derived from the RS saddle point equations \eq{RSsaddleQ1} and \eq{RSsaddleq1}. In the absence of the spin-lattice coupling $\delta=0$,
we find the spin and lattice degrees of freedom exhibit 'pseudo glass transitions' without RSB
as discussed already in sec.~\ref{sec:decoupled}. We show the decoupled phase diagrams in panels (a) and (b).
In the presence of the spin-lattice coupling $\delta > 0$ we find a new phase which we call
as 'spin-lattice glass phase' which accompanies RSB. As shown in panels (c),(d),(e), we always find the spin-lattice glass phase at low enough temperatures as long as $\delta >0$. In rigid regime
$\epsilon/J > 1$, the spin-lattice glass phase emerges within the pseudo lattice glass.
On the other hand, direct transition from the high temperature phase (paramagnetic and lattice liquid phase)
to the spin-lattice glass phase happens if the rigidity is low $\epsilon/J < 1$.

\subsection{Spin-lattice decoupled case $\delta=0$}\label{sec:decoupled}

In the absence of the spin-lattice coupling, i.e $\delta=0$, the free-energy
\eq{RSansatz-free-energy} as well as the saddle point equations  \eq{RSsaddleq1}
become decoupled into the spin part and lattice part.
As noted in sec~\ref{sec:decoupled}, the two decoupled systems are essentially the same
as the spherical SK model \cite{kosterlitz1976spherical}.

The pseudo lattice glass transition occurs at,
\beq
T^*(\epsilon) = \epsilon.
\eeq
and below $T^*$, $q>0$ solution is obtained as,
\beq
q = 1- T/\epsilon = t^*,
\label{eq-q-separated}
\eeq
where
\beq
t^* = (T^*-T)/T^*
\label{eq-t-star}
\eeq
is a dimensionless temperature that measures the distance of the temperature to the pseudo lattice glass transition
temperature $T^{*}$.
From the analysis of the Hessian matrix \eq{hessian}, we find that the $q > 0$ RS solution is marginally stable, i.e., the minimum eigenvalue of the Hessian matrix is strictly zero in the pseudo lattice glass phase.
Correspondingly, the glass susceptibility of the lattice distortion $\chi_{qq}$, given in \eq{chi_sg}, diverges approaching $T^*$ from above with the power law $|t^*|^{-1}$, and remain divergent within the lattice glass phase. The detailed analysis of the Hessian matrix and the glass susceptibility are presented in Appendix \ref{sec:stability} and Appendix. \ref{sec:sus}, respectively.

The same analysis can be repeated for the spin degrees of freedom and
one easily obtain the phase diagrams as shown in  Fig.~\ref{phase_mean} (a)(b).
In the following we turn to the cases of finite spin-lattice coupling $\delta >0$.

\subsection{High rigidity case} \label{sec:separated}

\begin{figure}[t]
\centering
\includegraphics[width=85mm]{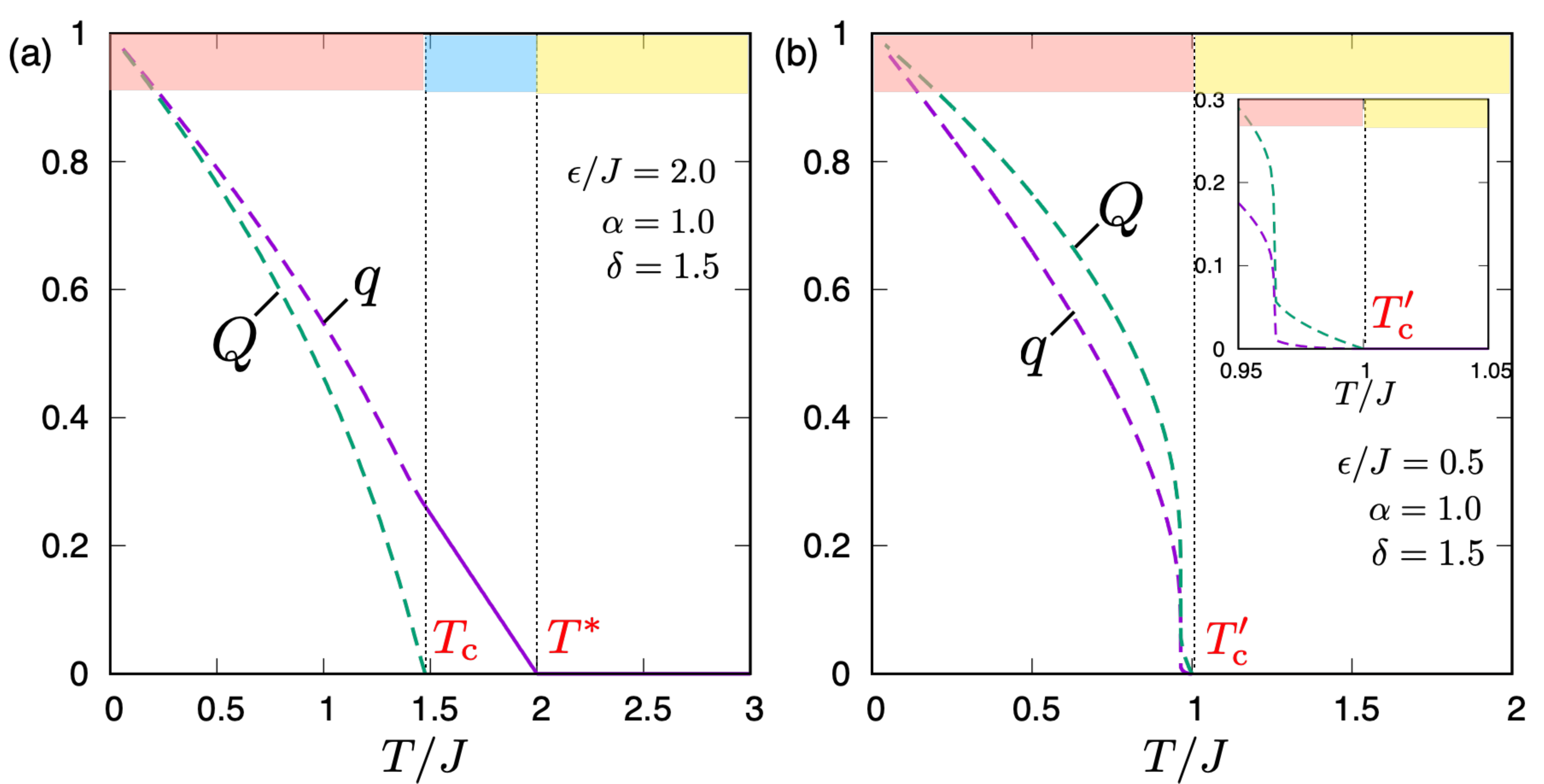}
\caption{
  Temperature dependencies of the order parameter $Q$ and $q$ in the RS ansatz for
  (a) high rigidity case ($\epsilon/J = 2, \alpha = 1, \delta = 1.5$)
  and (b) low rigidity case ($\epsilon/J = 0.5, \alpha = 1, \delta = 1.5$). The inset shows the behavior around $T_{\rm c}'$.
The broken lines represent the RS solution where it is unstable.
}
\label{orderparameter}
\end{figure}

Let us first discuss the case of high rigidity, i.~e. the lattice-lattice coupling is strong $\epsilon/J > 1$.
Lowering the temperature, we first pass over the pseudo lattice glass transition temperature $T^{*}(\epsilon)$
discussed above, which does not depend on the strength of the spin-lattice coupling $\delta$.
The lattice order parameter $q$ linearly grows lowering the temperature below $T^*$ 
while the spin glass order parameter $Q$ still remains zero at high enough temperatures,
as shown in Fig. \ref{orderparameter} (a). As shown in Fig. \ref{chi_sg} (a), $\chi_{qq}$ diverges passing $T^{*}$ 
while $\chi_{QQ}$ and $\chi_{Qq}$ do not exhibit any singularity as expected.
We also show in the figure the behaviour of the order parameters
below $T_{\rm c}$ using the RS solution in broken lines.  Note that the RS solution is unstable below $T_{\rm c}$ due to RSB.

By lowering the temperature further, the spin glass transition occurs at,
\beq
\frac{T_{\rm c}}{J} = \frac{J}{\epsilon}\qty(\sqrt{\delta^4 + \qty(\frac{\epsilon}{J})^2(2\delta^2 + 1)} - \delta^2).
\label{t_c_separated}
\eeq
below which $Q>0$ solution exist. 
It can be seen that the phase boundary given by \eq{t_c_separated} is shifted to higher temperatures by increasing $\delta$, and it converges to  $T_{\rm c} = T^* = \epsilon$ in the limit $\delta \to \infty$, i.e., the
pseudo lattice glass phase disappears in this limit.
 
 Below the spin glass transition temperature $T_{\rm c}$, the solutions of the order parameters are
 obtained from the saddle point equations
\eq{RSsaddleQ1} and \eq{RSsaddleq1} which can be written as,
\begin{align}
Q &= 1- \frac{T}{J\sqrt{1+2\delta^2 q}}, \label{saddlebelow1} \\
Q &= \frac{\sqrt{\alpha}T}{J\delta}\sqrt{\frac{q}{(1-q)^2} - \qty(\frac{\epsilon}{T})^2q}, \label{saddlebelow2}
\end{align}
The equations can be solved numerically as shown in Fig. \ref{orderparameter} (a).
Just below $T_{\rm c}$, the spin glass order parameter continuously increases as
\beq
Q = (1 + \beta_c J^2 \delta^2/\epsilon) t + \mathcal{O}(t^2), \label{RSQ2}
\eeq
where $\beta_c = 1/T_{\rm c}$ and
\beq
t = (T_{\rm c}-T)/T_{\rm c}
\label{eq-t}
\eeq
measures the distance of the temperature to the spin glass transition temperature $T_{\rm c}$.
On the other hand, the lattice glass order parameter is also modified reflecting the spin glass transition,
\begin{align}
q &= t^* + u, \\
u &= \qty(\frac{J \delta}{\epsilon})^2\frac{Q^2}{2 \alpha(\beta_c \epsilon -1)} \sim t^2 \label{RSq2}.
\end{align}

\begin{figure}[t]
\centering
\includegraphics[width=85mm]{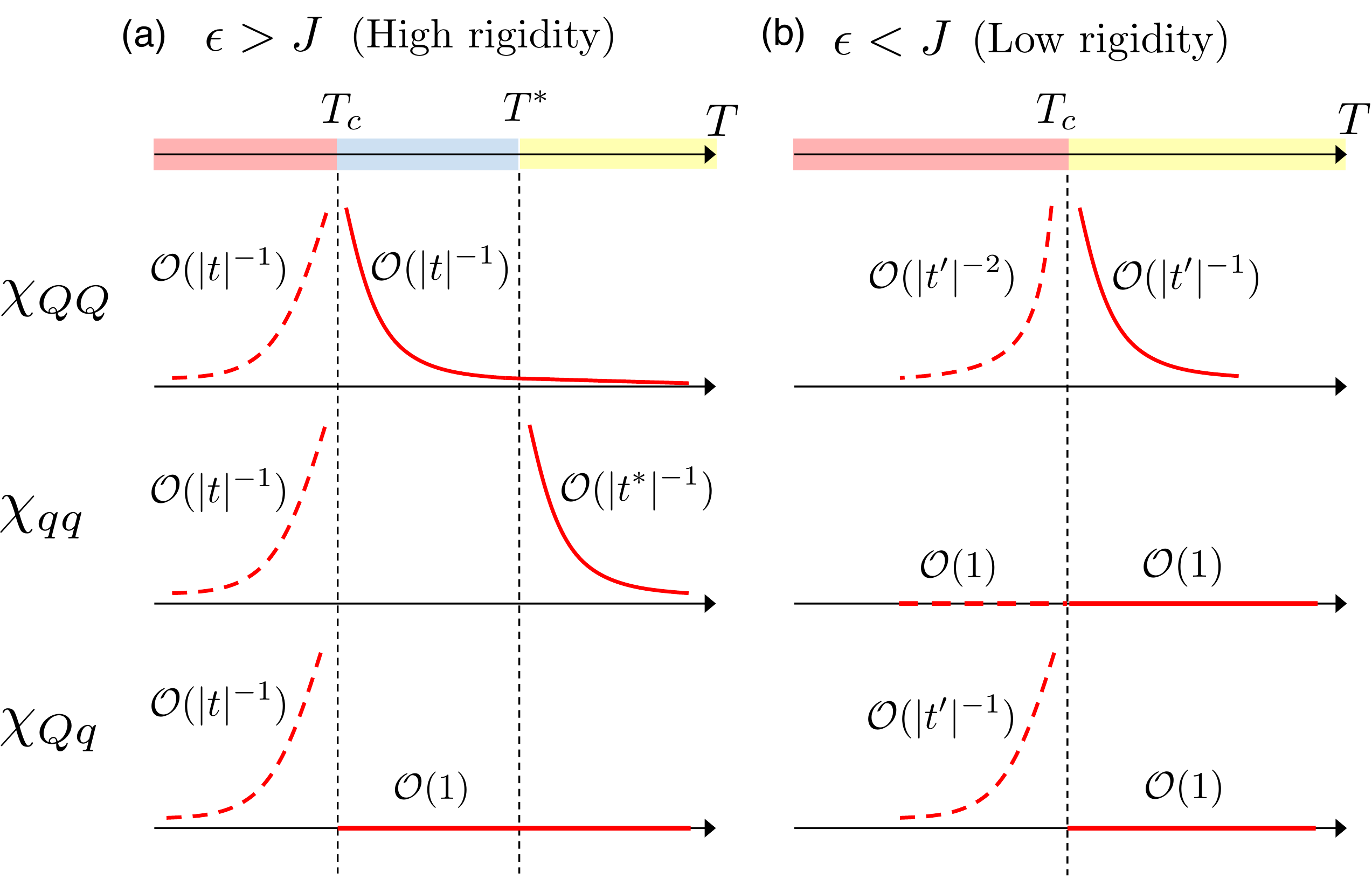}
\caption{
Schematic picture of the temperature dependence of the glass susceptibilities.
$t=(T_{\rm c}-T)/T_{\rm c}, t^*=(T^*-T)/T_{\rm c}$ and $t'=(T_{\rm c}'-T)/T_{\rm c}'$ are the dimensionless temperatures normalized by the glass transition temperatures. 
The broken lines represent the RS solution where it is unstable.
}
\label{chi_sg}
\end{figure}

Interestingly, the minimum eigenvalue of the Hessian matrix becomes negative below $T_{\rm c}$,
\beq
\lambda_{\rm min} = -2(\beta J \delta)^2 Q \sim -t.
\eeq
This means replica symmetry breaking must occur below $T_{\rm c}$.
Most probably full RSB occurs.
Note that the eigenvalue is quadratically proportional to the strength of spin-lattice coupling $\delta$.
Thus, spin-lattice coupling is essential for this model to exhibit the RSB.

Approaching $T_{\rm c}$ from above, the spin glass susceptibility $\chi_{QQ}$ diverges with power law $|t|^{-1}$.
Below  $T_{\rm c}$, the spin glass susceptibility $\chi_{QQ}$
as well as the lattice glass susceptibility decay with $|t|^{-1}$ within the RS ansatz.
The cross glass susceptibility $\chi_{Qq}$ doesn't grow above $T_{\rm c}$, but it suddenly diverges at $T_{\rm c}$ and decays with $|t|^{-1}$ below $T_{\rm c}$.

\begin{figure}[t]
\centering
\includegraphics[width=85mm]{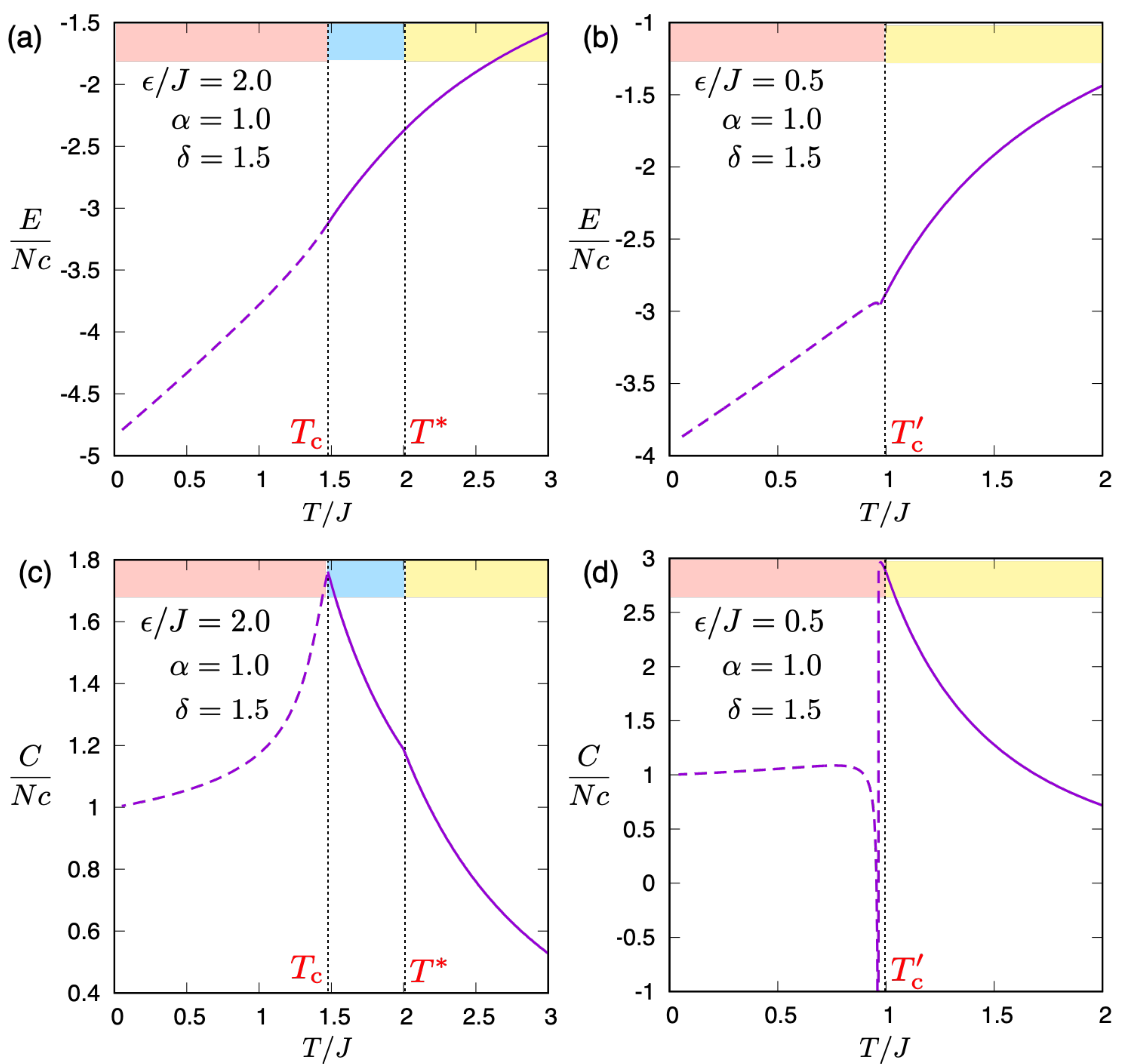}
\caption{(a), (b): Internal energy and heat capacity in the high rigidity case ($\epsilon/J = 2, \alpha = 1, \delta = 1.5$).
(c), (d): Internal energy and heat capacity in the low rigidity case ($\epsilon/J = 0.5, \alpha = 1, \delta = 1.5$).
$T^*,T_{\rm c}$ and $T_{\rm c}'$ are the glass transition temperatures obtained above. The broken lines represent the RS solution where it is unstable.
}
\label{energy_specific_mean}
\end{figure}

Fig.\ref{energy_specific_mean} (a) and (c) shows the internal energy and the heat capacity for $\epsilon/J = 2, \alpha = 1$ and $\delta = 1.5$ obtained by the RS solution (see  Appendix. \ref{sec:ene}).
Note that the heat capacity exhibits a kink at $T^*$ and a cusp at $T_{\rm c}$.

\subsection{Low rigidity case}\label{sec:simultaneous}

If the rigidity is low $\epsilon/J < 1$, the situation changes: a simultaneous spin and lattice glass transition takes place at a common transition temperature,
\beq
T_{\rm c}'/J =1,
\eeq
which doesn't depend on the elastic energy scale $\epsilon$ nor the amplitude $\delta$ of the spin-lattice coupling. The latter implies that the spin degrees of freedom play the dominant role in this case.

Below $T_{\rm c}'$, the non-zero order parameters $q,Q > 0$
are obtained by solving Eqs. (\ref{saddlebelow1}) and (\ref{saddlebelow2})
numerically as shown in Fig. \ref{orderparameter} (b).
We find continuous growth of the spin and lattice glass order parameters passing $T_{\rm c}'$ lowering the temperature. Note that both of the glass order parameters exhibit jumps slightly below $T_{\rm c}'$.
Using the dimensionless temperature,
\beq
t' = (T_{\rm c}' -T)/T_{\rm c}'
\label{eq-t-dash}
\eeq
which measures the distance of the temperature to $T_{\rm c}'$,
the order parameters $Q$ and $q$ slightly below $T_{\rm c}'$ can be obtained as,
\begin{align}
\nonumber
Q &= 1 - T/J + \frac{\delta^4}{\alpha(1 - (\epsilon/J)^2)}t'^2 + \mathcal{O}(t'^3) \\
&= t' + \delta^2 q  + \mathcal{O}(t'^3), \label{RSQ3} \\
q &= \frac{\delta^2}{\alpha(1 - (\epsilon/J)^2)}t'^2  + \mathcal{O}(t'^3). \label{RSq3}
\end{align}
Comparing the two order parameters, we notice that 
the growth of the lattice glass order parameter $q$ is weaker than that of the spin glass order parameter $Q$. 
More precisely, the critical exponents are different: $q$ is proportional to $t^2$ while $Q$ is proportional to $t$. If we switch off the spin-lattice coupling, $\delta \to 0$, the lattice glass order parameter $q$ becomes zero, and the behavior of the spin glass order parameter $Q$ agrees with the spherical SK model \cite{kosterlitz1976spherical}, i.e., $Q=t'$.
The spin glass order parameter $Q$ is weakly modified by the lattice glass order parameter $q$
at order $O(t'^{2})$.

From the analysis of the Hessian matrix, we find that the RS solution becomes unstable below $T_{\rm c}'$.
Slightly below $T_{\rm c}'$, the minimum eigenvalue behaves as,
\beq
\lambda_{\rm min} = -\frac{4(\beta J \delta)^4Q^2}{1-(\beta \epsilon)^2} \sim -(t')^2,
\eeq
which also becomes zero if $\delta = 0$.
The spin glass susceptibility $\chi_{QQ}$ exhibits a diverging feature approaching $T_{\rm c}$ from above with
power law $|t'|^{-1}$ while it decays as $|t|^{-2}$ below $T_{\rm c}$, reflecting $\lambda_{\rm min}$, as shown in Fig. \ref{chi_sg} (b).
In sharp contrast, the lattice glass susceptibility does not diverge at $T_{\rm c}$.
Similarly to the case of the separated glass transition $\epsilon/J < 1$, the cross glass susceptibility $\chi_{Qq}$ does not grow above $T_{\rm c}$, suddenly diverges at $T_{\rm c}$ and decays as $|t'|^{-1}$ below $T_{\rm c}'$.

Fig.\ref{energy_specific_mean} (b) and (d) shows the internal energy and the heat capacity for $\epsilon/J = 0.5, \alpha = 1$ and $\delta = 1.5$ using the RS solution.
Reflecting the jumps of the order parameters shown in Fig. \ref{orderparameter} (b), the energy exhibits a non-monotonic behavior, and the heat capacity diverges negatively.
We believe that the latter unphysical behavior is due to the failure of the RS solution below $T_{\rm c}'$.

\section{Discussion and Summary}\label{sec:summary}

In summary, we have constructed a disorder-free mean-field model of the spin-lattice coupled system on the tree-like $(k-1)$-simplices network. Using the replica method, we solved the model exactly in the dense limit $N \gg c \gg 1$ corresponding to the large dimensional limit $d \to \infty$. We have found the spin-lattice glass phase where both the spin and lattice degrees of freedom are frozen. The replica symmetry is broken there through the spin-lattice coupling yielding a complex and hierarchical free-energy landscape.

One may wander why a spin glass transition without the pseudo lattice glass transition does not happen
while the pseudo lattice glass transition without a spin glass transition happens at $T^{*}(\epsilon)$.
This is due to the spin-lattice coupling of the form $((\bm{\sigma}_i \cdot \hat{\bm{r}}_{ij}) \bm{S}_i \cdot \bm{S}_j + (\bm{\sigma}_j \cdot \hat{\bm{r}}_{ji}) \bm{S}_i \cdot \bm{S}_j)$ (see \eq{hami1b}).
When the lattice distortions become frozen, the lattice degrees of freedom play the role of bond randomness for the spins, while when the spins become frozen, the spin degrees of freedom play the role of random fields for the lattice distortions. The spin-glass transition does not take place until the effective coupling between spins 
become sufficiently strong. On the other hand, when a non-zero random field sets-in
it inevitably acts as a polarizing field.
Thus when the spins become frozen, the glass order parameter of the lattice distortions
necessarily becomes finite but not vice versa.

Now let us compare the results obtained in the mean-field model and the numerical observations
in the three-dimensional model on the pyrochlore lattice \cite{mitsumoto2020spin}.
In the low rigidity regime $\epsilon/J < 1$,
the mean-field model undergoes simultaneous spin and lattice glass transitions.
This point would appear to be consistent with the numerical observation in the
three-dimensional system \cite{mitsumoto2020spin}.
However the crucial difference is that in the mean-field case the glass susceptibility
associated with the lattice degrees of freedom does not exhibit a singularity in this regime.
This may be related to the fact that the mechanism to enable the non-zero lattice
glass order parameter, in this regime, is the random field effect mentioned above.

Next let us turn to the high rigidity regime $\epsilon/J >1$.
Here it is interesting to compare first the heat capacity.
In the three-dimensional model, the heat capacity exhibits two peaks at a high temperature and a low temperature
(see  Supplementary material Fig. S3 of Ref. \cite{mitsumoto2020spin}).
Interestingly the heat capacity of the present mean-field model also exhibits two anomalies at different temperatures
in the case of high rigidity $\epsilon/J >1$:
there is a kink at $T^{*}$ and peak at $T_{\rm c}$ (see Fig. \ref{energy_specific_mean} (c) ).
In the three-dimensional model, the peak of the heat capacity
at the higher temperature reflects a crossover from simple liquid at higher temperatures
to ice-like liquid at lower temperatures where the 2-in-2-out ice rule is satisfied nearly perfectly.
On the other hand, the peak at the lower temperature reflects a glass transition where both the spin and lattice degrees of exhibit a simultaneous glass transition accompanying diverging behavior of non-linear magnetic and dielectric susceptibilities associated with the spin and lattice degrees of freedom.
On the other hand, in the mean-field model, the two anomalies
in the heat capacity reflect the two separated glass transitions.
The (pseudo) lattice glass transition temperature $T^* = \epsilon$ does not depend on the amplitude of the spin-lattice coupling $\delta$, and it can occur without the spin-lattice coupling $\delta=0$.
We speculate that this (pseudo) lattice glass transition is an artifact of the mean-field model \cite{fisher1991pathologies} which is replaced by a smooth crossover in three-dimensions: $T^*$ is the crossover temperature
below which the lattice-lattice interactions become significant.
In the three-dimensional model, the heat capacity peak at the lower temperature shifts to the higher temperatures by increasing $\epsilon$, but it saturates at the higher $\epsilon$ region.
Very similar  $\epsilon$ dependence can be seen in the glass transition temperature
$T_{\rm c}$ of the mean-field theory (See Fig.~\ref{phase_mean}).
Indeed $T_{\rm c}$ given in \eq{t_c_separated} converges to a constant $T_{\rm c}/J \to \sqrt{(2\delta^2 + 1)}$
in $\epsilon \to \infty$ limit where lattice degrees of freedom should be completely frozen.
This is consistent with the fact that
in the usual spin glass models with quenched disorder, the spin glass transition is proportional to the amplitude of the quenched disorder \cite{saunders2007spin}. Based on the above observations, we speculate that the
mean-field model in the high rigidity regime $\epsilon/J >1$ compares better with the
numerical observations made in the three-dimensional system.

\section*{Acknowledgments}

We warmly thank Chisa Hotta for the collaborations closely related to this work and many useful discussions.
This work was supported by KANENHI (No. 19H01812) (No. 21K18146) from MEXT, Japan.

\appendix

\section{Derivation of the free energy functional}\label{sec:derivation}

In order to observe the glass transition with spontaneous replica symmetry breaking, we explicitly put the coupling between replicas into the Hamiltonian as\cite{parisi1989mechanism,monasson1995structural,MP99, yoshino2018disorder},
\begin{align}
\nonumber
\beta H &= \sum_{a = 1}^n \sum_{\tri} \beta V(\vec{\bm{S}}_{(\tri)}^a, \vec{\sigma}_{(\tri)}^a) \\
&- \sum_{a<b}\sum_i \qty[ \Lambda_{ab}  \sum_\mu (S^a)_i^\mu (S^b)_i^\mu + \lambda_{ab}  \sum_\nu (\sigma^a)_{i}^\nu (\sigma^b)_{i}^\nu],
\label{rsbhami}
\end{align}
and then study the behavior of the glass order parameter matrices of spins and lattice distortions under the coupling fields,
\begin{align}
Q_{ab} &= \lim_{\Lambda_{ab}, \lambda_{ab} \to 0} \lim_{N \to \infty} \frac{1}{NM} \sum_{i=1}^N \sum_{\mu = 1}^M \expval{(S^a)_i^\mu(S^b)_i^\mu}_{\Lambda,\lambda}, \label{orderQlambda} \\ 
q_{ab} &= \lim_{\Lambda_{ab}, \lambda_{ab} \to 0} \lim_{N \to \infty} \frac{1}{Nc} \sum_{i=1}^N \sum_{\nu = 1}^c \expval{(\sigma^a)_{i}^\nu (\sigma^b)_{i}^\nu}_{\Lambda,\lambda}. \label{orderqlambda}
\end{align}
Here $\expval{...}_{\Lambda,\lambda}$ represents the thermal average in the presence of the symmetry breaking fields $\lambda, \Lambda$, and these become the same as \eq{orderQ} and \eq{orderq} after taking the limit $\lambda, \Lambda \to 0$. Below we closely follow the steps used in \cite{yoshino2018disorder}.

To obtain a free-energy functional in terms of the order parameters, we perform the Legendre transformation.
This can be done in practice by making use of the following identities,
\begin{align}
1 &= \bint_{-\infty}^\infty dQ_{ab} \delta \qty(Q_{ab} - \frac{1}{NM} \sum_{i = 1}^N \sum_{\mu = 1}^M (S^a)_i^\mu(S^b)_i^\mu), \label{identity1} \\
1 &= \bint_{-\infty}^\infty dq_{ab} \delta \qty(q_{ab} - \frac{1}{Nc} \sum_{i = 1}^N \sum_{\nu = 1}^c (\sigma^a)_{i}^\nu (\sigma^b)_{i}^\nu), \label{identity2}
\end{align}
for $a<b$.

Then the traces of spin can be expressed formally as,
\begin{align}
\nonumber
&\prod_{a'=1}^n \text{Tr}_{\bm{S}^{a'}}[\cdots] = \frac{1}{2^n}\prod_{a\le b} \qty(NM \bint_{-i\infty}^{i\infty} \frac{d \Lambda_{ab}}{2\pi i} \bint_{-\infty}^\infty dQ_{ab}) \\
&\times \exp \qty[NM\qty( \frac{1}{2} \sum_{a,b=1}^n \Lambda_{ab} Q_{ab} + \log Z_\Lambda)] \prod_{i,\mu} \expval{\cdots}_{i,\mu},
\end{align}
where $Z_\Lambda = ((2\pi)^n/\text{det}(\hat{\Lambda}))^{1/2}$,
\begin{align}
\nonumber
\expval{\cdots}_{i,\mu} &= \frac{1}{Z_\Lambda}\prod_{a'=1}^n \qty( \int_{-\infty}^\infty d(S_i^{a'})^\mu) \\
&\times\exp \qty[{-\frac{1}{2}\sum_{a,b=1}^n \Lambda_{ab}(S_i^a)^\mu(S_i^b)^\mu}][\cdots],
\end{align}
$\Lambda_{ab}$ is Lagrange multiplier of \eq{trace1} and \eq{identity1}, and
$\prod_{i,\mu} \expval{\cdots}_{i,\mu}$ means averages for all sites $i$ and components $\mu$.
In the limit $N\to \infty$, we can use a saddle-point method to integrate out $\Lambda_{ab}$.
and the saddle point equation which determines the saddle point $\Lambda_{ab}^*(\hat{Q})$ is given by,
\beq
Q_{ab} = (\Lambda^*)_{ab}^{-1}
\eeq
and we find,
\beq
\prod_{a'=1}^n \text{Tr}_{\bm{S}^{a'}}[\cdots] \propto \prod_{a\le b} \qty(\int_{-\infty}^\infty dQ_{ab}) e^{Ncs_{\rm ent}^{(s)}[\hat{Q}]} \prod_{i, \mu} \expval{\cdots}_{i,\mu},
\label{trace_s}
\eeq
where $s_{\rm ent}^{(s)}[\hat{Q}]$ represents the entropic contribution of spins to the free energy, which is given by \eq{entQ}.

Similarly, the trace of lattice displacement is formally obtained as,
\begin{align}
\nonumber
&\prod_{a'=1}^n \text{Tr}_{\bm{\sigma}^{a'}}[\cdots] = \frac{1}{2^n}\prod_{a\le b} \qty(Nc \bint_{-i\infty}^{i\infty} \frac{d \lambda_{ab}}{2\pi i} \bint_{-\infty}^\infty dq_{ab}) \\
&\times\exp \qty[ Nc\qty( \frac{1}{2} \sum_{a,b=1}^n \lambda_{ab} q_{ab} + \log Z_\lambda)] \prod_{i, \nu} \expval{\cdots}_{i,\nu},
\end{align}
where $Z_\lambda = ((2\pi)^n/\text{det}(\hat{\lambda}))^{1/2}$,
\begin{align}
\nonumber
\expval{\cdots}_{i, \nu} &= \frac{1}{Z_\lambda}\prod_{a' = 1}^n \qty(\int_{-\infty}^\infty d (\sigma_i^{a'})^\nu ) \\
&\times \exp \qty[ -\frac{1}{2}\sum_{a,b=1}^n \lambda_{ab} (\sigma_i^a)^\nu (\sigma_i^b)^\nu][\cdots],
\end{align}
$\lambda_{ab}$ is the Lagrange multiplier of \eq{trace2} and \eq{identity2}, and $\prod_{i,\nu} \expval{\cdots}_{i,\nu}$ represents averages for all sites $i$ and components $\nu$.
In the limit $N \to \infty$, we use the saddle point method to integrate out $\lambda_{ab}$.
The saddle point equation is given by,
\begin{align}
q_{ab} = (\lambda^*)_{ab}^{-1}.
\end{align}
We obtain,
\beq
\prod_{a'=1}^n \text{Tr}_{\bm{\sigma}^{a'}}[\cdots] \propto \prod_{a\le b} \qty(\int_{-\infty}^\infty dq_{ab}) e^{Ncs_{\rm ent}^{(\sigma)}[\hat{q}]} \prod_{i,\nu} \expval{\cdots}_{i,\nu},
\label{trace_sig}
\eeq
where $s_{\rm ent}^{(\sigma)}[\hat{q}]$ represents the entropic contribution of spins to the free energy, which is given by \eq{entq}.

Supposing the translational and rotational symmetry, where all sites $i$ and components $\mu,\nu$ are equivalent to each other, we find
\begin{align}
\label{useful}
\begin{aligned}
&\prod_{i,\mu} \expval{(S_i^a)^{\mu'}}_{i,\mu} = 0, \\
&\prod_{i,\mu} \expval{(S_i^a)^{\mu'} (S_i^b)^{\mu'}}_{i,\mu} = Q_{ab}, \\
&\prod_{i,\nu} \expval{(\sigma_i^a)^{\nu'}}_{i,\nu} = 0, \\
&\prod_{i,\nu} \expval{(\sigma_i^a)^{\nu'} (\sigma^b)^{\nu'}}_{i,\nu} = q_{ab}.
\end{aligned}
\end{align}
These relations are very useful for performing the cumulant expansion, and we can obtain the exact form of the replicated free energy. 
Now the replicated partition function can be rewritten as,
\begin{align}
\nonumber
Z^n &\propto \prod_{a \le b}\qty(\int_{-\infty}^\infty dQ_{ab} \int_{-\infty}^\infty dq_{ab}) \\
&\times \exp \qty[Nc \qty(s_{\rm ent}^{(s)}[\hat{Q}]+ s_{\rm ent}^{(\sigma)}[\hat{q}]+\mathcal{F}_{\rm int}[\hat{Q},\hat{q}])],
\end{align}
where
\begin{align}
\nonumber
&Nc\mathcal{F}_{\rm int} \\ \nonumber
= &\log \qty(\prod_{i,\mu,\nu}  \expval{e^{-\sum_{a=1}^n \sum_{\tri=1}^{N_{\tri}} \beta V(\vec{\bm{S}}^a_{(\tri)}, \vec{\sigma}^a_{(\tri)})}}_{i, \mu, \nu}) \\
= &\log \qty ( \prod_{i,\mu,\nu} \expval{
e^{\frac{-\beta}{\sqrt{c(k-1)}} \sum_{a=1}^n\sum_{\tri=1}^{N_{\tri}}\sum_{i<j}^k A_{ij(\tri)}^a} }_{i, \mu, \nu} ).
\end{align}
with
\begin{align}
\nonumber
A_{ij}^a &= J \qty(1+\delta(\bm{\sigma}^a_i \cdot \hat{\bm{r}}_{ij} + \bm{\sigma}^a_j \cdot \hat{\bm{r}}_{ji})) \sum_{\mu = 1}^M (S_{i}^a)^\mu (S_{j}^a)^\mu \\
&+ \epsilon \sum_{\nu = 1}^c (\sigma_{i}^a)^\nu (\sigma_{j}^a)^\nu.
\end{align}
$\prod_{i,\mu,\nu}\expval{\cdots}_{i,\mu,\nu}$ means averages for all $i,\mu,\nu$.
Noting that the average can be decoupled with respect to different site $i~(\tri)$, spin component $\mu$, and distortion component $\nu$, 
we can evaluate $\mathcal{F}_{\rm int}$ using the cumulant expansion,
\begin{align}
\nonumber
&\log \qty(\prod_{i,\mu,\nu} \expval{e^{\frac{-\beta}{\sqrt{c(k-1)}}\sum_{a}\sum_{\tri}\sum_{i<j}A_{ij(\tri)}^a}}_{i, \mu, \nu}) \\ \nonumber 
&= \frac{-\beta}{\sqrt{c(k-1)}} N_\tri \mqty(k\\2) \sum_a \expval{A_{ij}^a} \\
&+ \frac{1}{2!} \frac{\beta^2}{c(k-1)} N_\tri \mqty(k\\2) \sum_{a,b} \qty( \expval{A_{ij}^a A_{ij}^b} - \expval{A_{ij}^a}\expval{A_{ij}^b}) + \cdots
\end{align}
with
\begin{align}
\expval{A_{ij}^a} &= 0, \\
\expval{A_{ij}^aA_{ij}^b} &= J^2 \qty(1+2\delta^2q_{ab})Q_{ab}^2+ \epsilon^2 q_{ab}^2.
\end{align}
Here, we used $\sum_{\nu = 1}^c (\hat{r}_{ij}^\nu)^2 = 1$ and \eq{useful}.
Higher order cumulants vanish in the dense limit\cite{yoshino2018disorder}, i.e. $\lim_{c \to \infty}$ after $\lim_{N \to \infty}$. Finally, we obtain \eq{Fint} using \eq{eq-N_tri}.

\section{Stability of the RS solution}\label{sec:stability}

Now, let us investigate the stability of the RS solution, which is associated with the divergence of the glass susceptibility, shown in Appendix. \ref{sec:sus}.
The necessary condition of the RS solution to be stable is that the eigenvalues of Hessian matrix $\mathcal{H}$ given by,
\beq
\hat{\mathcal{H}} = \mqty[H_{QQ}&H_{Qq}\\H_{qQ}&H_{qq}],
\eeq 
are all non-negative, where the submatrices $H_{QQ}$, $H_{Qq} = H_{qQ}$ and $H_{qq}$ are given by \eq{hessian}.
To analyze these matrices we need the first and second derivatives of the functional $s_n[\hat{Q},\hat{q}]$, which are obtained as,
\begin{align}
\pdv{s_n}{Q_{ab}}{Q_{cd}} &= -\frac{1}{\alpha}\qty(Q_{ac}^{-1}Q_{bd}^{-1} + Q_{ad}^{-1}Q_{bc}^{-1}) \nonumber \\ 
&+ \delta_{ac}\delta_{bd} \frac{(\beta J)^2}{\alpha}(1+2\delta^2 q_{ab}) \\
\pdv{s_n}{Q_{ab}}{q_{cd}} &=  2\delta_{ac}\delta_{bd} (\beta J \delta)^2 Q_{ab} \\
\pdv{s_n}{q_{ab}}{q_{cd}} &=  \qty[ -\qty(q_{ac}^{-1}q_{bd}^{-1} + q_{ad}^{-1}q_{bc}^{-1}) + \delta_{ac}\delta_{bd} (\beta \epsilon)^2],
\end{align}
where $a<b, c<d$. 
In the replica symmetric case we have $Q_{ab}=(1-Q)\delta_{ab}$ and $q_{ab}=(1-q)\delta_{ab}+q$
(see  \eq{RSansatz}).  
Then submatrices of the Hessian matrix in the limit $n \to 0$ can be cast into the following form,
\begin{align}
\qty{H_{QQ}}_{ab,cd} &= M_1^Q\frac{\delta_{ac} \delta_{bd} + \delta_{ad} \delta_{bc}}{2} \nonumber \\
&+ M_2^Q\frac{\delta_{ac} + \delta_{ad} + \delta_{bc} + \delta_{bd}}{4} + M_3^Q \label{hesseQQ} \\
\qty{H_{Qq}}_{ab,cd} &=  \gamma \delta_{ac}\delta_{bd} \\
\qty{H_{qq}}_{ab,cd} &= M_1^q\frac{\delta_{ac} \delta_{bd} + \delta_{ad} \delta_{bc}}{2} \nonumber \\
&+ M_2^q\frac{\delta_{ac} + \delta_{ad} + \delta_{bc} + \delta_{bd}}{4} + M_3^q, \label{hesseqq}
\end{align}
where
\begin{eqnarray}
\nonumber
  M_1^Q &=& \frac{2}{\alpha(1-Q)^2} - \frac{2(\beta J)^2(1 + 2\delta^2 q)}{\alpha}   \\
  M_1^q &=& \frac{2}{(1-q)^2} - 2(\beta \epsilon)^2\\
  M_2^Q & =& \frac{-4Q}{\alpha(1-Q)^3} \qquad
  M_2^q  = \frac{-4q}{(1-q)^3}\\
M_3^Q &=& \frac{-2Q^2}{\alpha(1-Q)^4} \qquad
M_3^q = \frac{-2 q^2}{(1-q)^4}
\end{eqnarray}
and
\beq
\gamma = -2(\beta J \delta)^2 Q.
\label{c_Q}
\eeq

First let us analyze the eigenvalues of the submatrices $H_{QQ}$ and $H_{qq}$ before considering the eigenvalues of the total Hessian matrix $\hat{\mathcal{H}}$.
From Eqs. (\ref{hesseQQ}) and (\ref{hesseqq}), we find that both $H_{QQ}$ and $H_{qq}$ become diagonalized by the same orthogonal matrix $P$ and then the eigenvalues $\lambda_k^\omega$ for $k=1,2,...,n(n-1)/2$ and $\omega = Q, q$ are obtained as 
\beq
\lambda_k^\omega = 
\begin{cases}
\lambda_L^\omega ~~~(k = 1) \\
\lambda_A^\omega ~~~(2 \le k \le n) \\
\lambda_R^\omega ~~~(n+1 \le k \le n(n-1)/2),
\end{cases}
\label{label_k}
\eeq
with
\begin{align}
\lambda_L^\omega &= \frac{1}{2}M_1^\omega + \frac{(n-1)}{2}M_2^\omega + \frac{n(n-1)}{2}M_3^\omega \nonumber \\
&\xrightarrow[n \to 0]{} \frac{1}{2}(M_1^\omega - M_2^\omega), \\
\lambda_A^\omega &= \frac{1}{2}M_1^\omega + \frac{n-2}{4}M_2^\omega \xrightarrow[n \to 0]{} \frac{1}{2}(M_1^\omega - M_2^\omega), \\
\lambda_R^\omega &= \frac{1}{2}M_1^\omega.
\end{align}
$\lambda_R$, called as the replicon mode, is the minimum one among the three eigenvalues if the order parameters $Q$ and $q$ are positive.
Therefore replica symmetry breaking occurs when the replicon mode $\lambda_R$ becomes negative.

Now let us introduce a matrix $\hat{\Xi}$ of size $n(n-1) \times n(n-1)$,
\begin{align}
\hat{\Xi} \equiv 
&\mqty[P^{-1}&O\\O&P^{-1}]\mqty[H_{QQ}&H_{Qq}\\H_{qQ}&H_{qq}]\mqty[P&O\\O&P] \nonumber \\ &= \mqty[ P^{-1}H_{QQ}P & \gamma\hat{I}_{n(n-1)/2} \\ \gamma\hat{I}_{n(n-1)/2}  & P^{-1}H_{qq}P]
\label{Xi}
\end{align}
where $\hat{I}_{m}$ is identity matrix of size $m \times m$. Note that the diagonal submatricies
$P^{-1}H_{QQ}P$ and $P^{-1}H_{qq}P$ are already diagonalized within themselves.
The eigen values $\lambda$ of the total Hessian matrix $\hat{\mathcal{H}}$ must satisfy
the following equation
\beq
\det \qty[\hat{\Xi} - \lambda \hat{I}_{n(n-1)} ] = \prod_k \qty[(\lambda_k^Q-\lambda)(\lambda_k^q-\lambda)-\gamma^2] = 0
\eeq
which yields the eigen values,
\beq
\lambda_{k\pm} = \frac{\lambda_k^Q + \lambda_k^q \pm \sqrt{(\lambda_k^Q - \lambda_k^q)^2 + 4\gamma^2}}{2}\label{eigen} 
\eeq
for $k=1,2,..., n(n-1)/2$.
Note that $\lambda_{k\pm} = \lambda_k^Q,\lambda_k^q$ if $\gamma = 0$, which means that each degree of freedom doesn't affect the stability of another degree of freedom.
Thus, in the paramagnetic phase $Q=q=0$ (see Fig. \ref{phase_mean}), all eigenvalues are positive, 
\begin{align}
\lambda_{L-} = \lambda_{A-} = \lambda_{R-} = 
\begin{cases}
1-(\beta \epsilon)^2~(\epsilon > J), \\
\frac{1 - (\beta J)^2}{\alpha}~~(\epsilon < J).
\end{cases}
\label{para_eigen}
\end{align}
The $Q=q=0$ solution becomes unstable below the lattice glass transition temperature $T^*=\epsilon$
and the simultaneous glass transition temperature $T_{\rm c}'=J$.

\subsection{Spin-lattice decoupled case $\delta=0$}\label{sec-stability-decoupled}
In the absence of spin-lattice coupling, the two decoupled systems are essentially the same as the spherical SK model.
In the pseudo spin glass and pseudo lattice glass phases, the minimum eigenvalues become
\begin{align}
\lambda_R^Q &= \frac{M_1^Q}{2} =0, \\
\lambda_R^q &= \frac{M_1^q}{2} = 0,
\end{align}
respectively, i. e., marginally stable.
\subsection{High rigidity case}
\label{sec-stability-separated}

For $\epsilon >  J $, the lattice glass transition occurs at the transition temperature $T^*=\epsilon$
where $\lambda_{k-}$ for all $k$ become 0.
below $T^*$, i.e. in the lattice glass phase  with
$q=1-T/\epsilon$ (see \eq{eq-q-separated}), $\lambda_{L-} = \lambda_{A-}$ becomes positive again,
\beq
\lambda_{\rm L-} = \lambda_{\rm A-} = \frac{2 q}{(1-q)^3} >0
\eeq
while the replicon mode remains marginal,
\beq
\lambda_{R-}=\frac{M_{1}^{q}}{2}=0
\eeq
It means that the system becomes marginally stable against RSB
in the lattice glass phase similar to the spherical spin glass model \cite{kosterlitz1976spherical}.
There is another glass transition temperature $T_{\rm c} (< T^{*})$ \eq{t_c_separated}below which $Q>0$ solution emerges. One can check that $M_{1}^{Q} >0$ for $Q=0$ above $T_{\rm c}$ but vanishes at $T_{\rm c}$.
where 
we can evaluate the replicon mode as,
\beq
\lambda_{R-} =\frac{M_{1}^{Q}}{2}= \gamma + O(t^2) \sim -t. 
\eeq
where $t$ is the distance of the temperature to $T_{\rm c}$ (see \eq{eq-t}).
Therefore, the replica symmetry should be broken below $T_{\rm c}$.

\subsection{Low rigidity case}
\label{sec-stability-simultaneous}

For $\epsilon < J $, the simultaneous spin and lattice glass transition occurs at $T_{\rm c}'$ and $\lambda_{k-}$ for all $k$ becomes 0.
Below $T_{\rm c}'$, $\lambda_{A-} = \lambda_{L-}$ becomes positive again,
\beq
\lambda_{A-} = \lambda_{L-} = \frac{2 Q}{\alpha(1-Q)^3},
\eeq
and the replicon mode becomes 
\begin{align}
\nonumber
\lambda_{R-} = -\frac{\gamma^2}{\lambda_R^q} = -\frac{4(\beta J \delta)^4Q^2}{1-(\beta \epsilon)^2} \sim -t^{'2}.
\end{align}
Therefore, the continuous replica symmetry breaking takes place at $T_{\rm c}'$.

\section{Glass susceptibility}\label{sec:sus}

The glass susceptibility is given by the inverse matrix of the Hessian matrix $\hat{\mathcal{H}}$, i.e. 
\beq
\hat{\chi} = \mqty[\chi_{QQ} & \chi_{Qq} \\ \chi_{qQ} & \chi_{qq}] = \mqty[H_{QQ} & H_{Qq} \\ H_{qQ} & H_{qq}]^{-1}.
\eeq
In order to obtain the $n(n-1) \times n(n-1)$ components $\chi_{ij}$, let us consider the matrix $\hat{S}$ defined such that
\begin{align}
\hat{S}^{-1} \hat{\mathcal{H}} \hat{S} &=  \mqty[\lambda_{1+}&&&&&\\ &\ddots&&&&\\&&\lambda_{m+}&&&\\&&&\lambda_{1-}&&\\ &&&&\ddots&\\&&&&&\lambda_{m-}], 
\end{align}
where $m = n(n-1)/2$.
Using matrix $\hat{S}$, we find,
\beq
\chi_{ij} = \sum_{l} \frac{s_{il}s_{jl}}{\lambda_l}
\eeq
where $s_{ij}$ is the components of the matrix and $\lambda_l$ is a short hand notation defined such that
\beq
\lambda_l = 
\begin{cases}
\lambda_{l+} ~~~~~~~~(l \le m) \\
\lambda_{(l-m)-} ~~~(m < l).
\end{cases}
\eeq
Using \eq{Xi}, we notice that $\hat{S}$ can be factorized as
\beq
\hat{S} = \mqty[P&O\\O&P]\hat{U},
\eeq
where $\hat{U}$ is a matrix which diagonalize $\hat{\Xi}$ (\eq{Xi}), i.e.
\beq
\hat{U}^{-1}\hat{\Xi}\hat{U} =  \mqty[\lambda_{1+}&&&&&\\ &\ddots&&&&\\&&\lambda_{m+}&&&\\&&&\lambda_{1-}&&\\ &&&&\ddots&\\&&&&&\lambda_{m-}],
\eeq
where $\lambda_k^Q$ and $\lambda_k^q$ are given by \eq{label_k}.
Here, each eigenvector is normalized as
\beq
\sum_{i=1}^{2m}u_{ij}^2 = 1 ~~\text{for all $j$},
\eeq
where $u_{ij}$ is the component of the matrix $\hat{U}$.
If $\gamma = 0$, i.e. in the paramagnetic phase ($Q=q=0$) or the pseudo lattice glass phase ($Q=0,q>0$), $\hat{U}$ becomes an identity matrix $\hat{I}_{2m}$.
In the spin-lattice glass phase ($Q,q>0$), we obtain the elements of the matrix $\hat{U}$ as,
\beq
u_{kj} =
\begin{dcases}
u_{(k+m)j} = 0 ~~~~~~~~~~\text{if }~\lambda_j = \lambda_{k\pm}  \\
 - \frac{\lambda_k^q - \lambda_j}{\gamma}u_{(k+m)j}~~~\text{if } ~ \lambda_j \neq \lambda_{k\pm}
\end{dcases}
\label{t_compponent}
\eeq
for $k = 1,2,...,m$.

Now we find the inverse matrix of $\hat{\Xi}$ is obtained as,
\beq
\qty{\hat{\Xi}^{-1}}_{ij} = \sum_{l}\frac{u_{il}u_{jl}}{\lambda_l},
\label{xi_inverse_component}
\eeq
Then we obtain the glass susceptibility as,
\begin{align}
\nonumber
\hat{\chi} =\hat{\mathcal{H}}^{-1} &=
\hat{S} \mqty[1/\lambda_{1+}&&&&&\\ &\ddots&&&&\\&&1/\lambda_{m+}&&&\\&&&1/\lambda_{1-}&&\\ &&&&\ddots&\\&&&&&1/\lambda_{m-}]\hat{S}^{-1} \\
&=
\mqty[P&O\\O&P] \hat{\Xi}^{-1} \mqty[P^{-1}&O\\O&P^{-1}]
\end{align}
Note that all components of $P$ are $O(1)$.
Here, it is convenient to decompose $\hat{\Xi}^{-1}$ as,
\beq
\hat{\Xi}^{-1} = \mqty[\Xi^{-1}_{QQ}&\Xi^{-1}_{Qq}\\\Xi^{-1}_{qQ}&\Xi^{-1}_{qq}],
\eeq
since the components of $\chi_{\alpha \beta}~(\alpha, \beta = Q,q)$ are associated with only the components of $\Xi_{\alpha \beta}$.

We show the schematic picture of the temperature dependences of $\chi_{\alpha \beta}~(\alpha, \beta = Q,q)$ in Fig. \ref{chi_sg}.
In the following, we present the detail of the computation to obtain this.

\subsection{Spin-lattice decoupled case $\delta=0$}\label{sus_decoupled}

In the absence of the spin-lattice coupling, the matrix $\hat{U}$ becomes an identity matrix $\hat{I}_{2m}$.
Thus, the components of $\chi_{QQ}$ and $\chi_{qq}$ are proportional to $1/\lambda_k^{Q}$ and $1/\lambda_k^{q}$, respectively, which grow as $1/|t^*|$ above $T^*$ and diverge at $T^*$.
In the pseudo glass phase, the replicon modes are always zero, i.e., the glass susceptibilities always diverge.

\subsection{High rigidity case}\label{sus_sepa}

For $\epsilon/J >  1$, the system exhibits the lattice glass transition at $T^*$ and the spin glass transition at $T_{\rm c}$ which is lower than $T^*$ (see Fig. \ref{phase_mean}).
Slightly above $T^*$ the system is in the paramagnetic phase ($Q=q=0$) and $1/\lambda_k^q$ for all $k$ proportional to $1/|t^*|$ as discussed in sec.~\ref{sec-stability-separated}.
Thus we find all components in the sector $\chi_{qq}$ diverge at $T^*$, 
\beq
\qty{\chi_{qq}}_{ij} \sim O(|t^*|^{-1})~~~~(T>T^*),
\eeq
while $\chi_{QQ}$ and $\chi_{Qq}$ have no singularity since the spin-lattice coupling does not work
as long as $\gamma \propto Q=0$.

Slightly above $T_{\rm c}$ the system is in the lattice glass phase ($Q=0,q\neq0$) and $1/\lambda_k^Q$ for all $k$ grows as $\propto 1/|t|$ as $t \to 0$.
The components in the sector  $\chi_{QQ}$ now diverge at $T_{\rm c}$, 
\beq
\qty{\chi_{QQ}}_{ij} \sim O(|t|^{-1})~~~~(T>T_{\rm c}),
\eeq
while $\chi_{Qq}$ still has no singularity since $Q=0$.

In the spin-lattice glass phase, now $\gamma$ becomes finite so that $\hat{U}$ is no more an identity matrix.
Let us consider the components of $\hat{\Xi}^{-1}$ related to $\lambda_{R\pm}$ because they are relevant to the divergent features of the components of $\hat{\chi}$.
Using \eq{t_compponent} and \eq{xi_inverse_component} and noting $(\lambda_{R}^q-\lambda_{R\pm})/\gamma \sim O(1)$, we find the magnitude of the components of $\hat{U}$ are,
\beq
\hat{U} =  \mqty[O(1)&O(1) \\ O(1)&O(1)].
\eeq
Here each block size is $m \times m$.
Since $\frac{1}{\lambda_{R-}} \sim |t|^{-1}$, the components of $\hat{\Xi}^{-1}_{ij}$ become
\beq
\qty{\Xi^{-1}_{QQ}}_{ij} \sim \qty{\Xi^{-1}_{Qq}}_{ij} \sim \qty{\Xi^{-1}_{qq}}_{ij} \sim O(|t|^{-1}).
\eeq
Therefore, the components of $\hat{\chi}$ slightly below $T_{\rm c}$ are obtained as
\beq
\qty{\chi_{QQ}}_{ij} \sim \qty{\chi_{Qq}}_{ij} \sim \qty{\chi_{qq}}_{ij} \sim O(|t|^{-1})~~~~(T<T_{\rm c}).
\eeq

\subsection{Low rigidity case}\label{sus_simul}

For $\epsilon/J < 1$, the system exhibits the simultaneous spin and lattice glass transitions at $T_{\rm c}'$ (see Fig. \ref{phase_mean}).
Slightly above $T_{\rm c}'$ the system is in the paramagnetic phase ($Q=q=0$) and $1/\lambda_k^Q$ for all $k$ proportional to $1/|t'|$.
The components in the sector $\chi_{QQ}$ diverge at $T_{\rm c}'$,
\beq
\qty{\chi_{QQ}}_{ij} \sim O(|t'|^{-1})~~~~(T>T_{\rm c}'),
\eeq
while $\chi_{Qq}$ and $\chi_{qq}$ remain finite.

Similarly to the high rigidity case, in the spin-lattice glass phase, we consider the components of $\hat{\Xi}^{-1}$ related to the only $\lambda_{R\pm}$.
Using \eq{t_compponent} and \eq{xi_inverse_component} and noting $(\lambda_{R}^q-\lambda_{R-})/\gamma \sim O(|t'|^{-1})$ and $(\lambda_{R}^q-\lambda_{R+})/\gamma \sim O(|t'|)$, we find the magnitude of the components of $\hat{U}$ are evaluated as,
\beq
\hat{U} =  \mqty[O(|t'|)&O(1) \\ O(1)&O(|t'|)].
\eeq
Since $1/\lambda_{R-} \sim |t|^{-2}$, the components of $\hat{\Xi}^{-1}_{ij}$ become
\begin{align}
\qty{\Xi^{-1}_{QQ}}_{ij} &\sim O(|t'|^{-2}), \\
\qty{\Xi^{-1}_{Qq}}_{ij} &\sim O(|t'|^{-1}), \\
\qty{\Xi^{-1}_{qq}}_{ij} &\sim O(1).
\end{align}
Therefore, the components of $\hat{\chi}$ slightly below $T_{\rm c}$ are obtained as
\begin{align}
\qty{\chi_{QQ}}_{ij} &\sim O(|t'|^{-2}), \\
\qty{\chi_{Qq}}_{ij} &\sim O(|t'|^{-1}), \\
\qty{\chi_{qq}}_{ij} &\sim O(1).
\end{align}
The notable fact is that the glass susceptibility of lattice distortions $\qty{\chi_{qq}}_{ij}$ does not diverge by approaching the simultaneous glass transition temperature $T_{\rm c}'$ from both sides due to the pre-factor $u_{ij}$ while the lattice glass order parameter $q$ becomes finite.

\section{Energy, Heat capacity}\label{sec:ene}

The internal energy and heat capacity within the RS ansatz are given by,
\begin{align}
\nonumber
\frac{\expval{E}}{Nc} &= \frac{1}{Nc}\pdv{(\beta f^{\rm RS})}{\beta} \\ \nonumber
&= - \frac{\beta}{2} \Biggl\{ \frac{J^2}{\alpha}(1+2\delta^2) + \epsilon^2 \\
&- \qty[\frac{J^2}{\alpha}(1+2\delta^2 q)Q^2 + \epsilon^2q^2]\Biggr\}, \\
\frac{C}{Nc} &= \frac{1}{Nc} \frac{d \expval{E}}{dT}.
\end{align}

\bibliography{meanfield_yoshino}

\end{document}